\documentclass{optica-article}

\journal{opticajournal} 

\articletype{Research Article}

\usepackage{lineno}

\usepackage{array}
\usepackage{threeparttable}

\usepackage{subcaption}
\usepackage{tikz}
\usetikzlibrary{
    shapes.geometric,       
    arrows.meta,            
    positioning,            
    calc,                   
    backgrounds,            
    fit,                    
    decorations.pathreplacing, 
    matrix                  
}

\usepackage{algorithm}
\usepackage{algpseudocode}
\usepackage{siunitx}
\usepackage{hyperref}
\begin{document}

\title{Reinforcement learning enabled turnkey soliton generation in a 4H-SiC microresonator}


\author{Viswanathan Sankar,\authormark{1,\dag,*} Alice Houard,\authormark{2,\dag} Yongsheng Wang,\authormark{1} Paolo Leonelli,\authormark{1} Haiyan Ou,\authormark{1} Arnaud Mussot, \authormark{1} Matteo Conforti ,\authormark{2} Christophe Peucheret,\authormark{3} and Darko Zibar\authormark{1}}

\address{\authormark{1}DTU Electro, Technical University of Denmark, DK-2800 Kgs. Lyngby, Denmark\\
\authormark{2}Université de Lille, CNRS, UMR 8523-PhLAM-Physique des Lasers Atomes et Molécules, F-59000 Lille, France\\
\authormark{3}Univ. Rennes, CNRS, FOTON - UMR6082, 22305 Lannion, France}

\authormark{\dag}These authors contributed equally to this work.
\authormark{*}Corresponding author: viswa@dtu.dk

\begin{abstract*}

Practical deployment of microresonator frequency combs requires turnkey soliton generation without heavy computational overhead or bulky auxiliary hardware. We demonstrate a computationally inexpensive method combining reinforcement learning and PID control for single-soliton generation and retention in a microring resonator. It requires pump frequency tuning and two photodiodes, one monitoring the cavity transmission and the other behind an optical band-pass filter that selects the comb's long-wavelength tail. Rather than sweeping the pump frequency at a fixed rate, the agent chooses each tuning step from the measured photodiode signals, learning by trial and error how fast and in which direction to step as the intra-cavity field approaches the soliton. The agent captures a single soliton in 1.32\,s on average without operator's intervention, whereas manual tuning relies on spectrum-analyzer traces that take seconds to acquire for each tuning step. Without an explicit model of the intra-cavity dynamics, it succeeds in 98\% of 100 trials, compared with 78\% for the best of three fixed linear sweeps and 82\% for a scan whose slope is tuned after every sweep. Once the soliton is captured, a PID loop retains it, and if the comb is lost the agent resumes the search and reacquires it without operator's intervention, pointing toward turnkey, portable microcomb sources.
\end{abstract*}

\section{Introduction}
\label{sec:intro}

Dissipative Kerr soliton (DKS) in integrated high-Q microresonators offers compact chip-scale optical frequency comb sources \cite{diddams2020optical, pasquazi2018microcombs}. Due to their low phase noise, wide bandwidth and low power requirements, microcombs are particularly attractive for applications such as spectroscopy, ranging, optical telecommunications, and low-noise microwave generation \cite{herr2013si, sun2023review, corcoran2025optical}.
Practical deployment depends on the deterministic generation and stabilization of the soliton state. The standard approach is to tune the pump frequency across the cavity resonance to trigger a cavity soliton. This is complicated by photothermal absorption in the small mode volume, where concentrated power dissipation shifts the resonance strongly and produces dynamic hysteresis \cite{carmon2004dynamical}. As the pump is tuned into the resonance, the system first traverses a chaotic modulation-instability (MI) regime with high intracavity power. The transition to the soliton state is marked by an abrupt drop in intracavity power, which removes much of the thermal load. The resonator therefore cools and its resonance shifts rapidly away from the pump, increasing the effective pump-cavity detuning. If this detuning exceeds the limits of soliton existence range (SER), the soliton collapses to the continuous-wave (CW) state \cite{bao2017direct, yi2016active}. Because of these tight dynamic constraints, a conventional frequency sweep must be tuned to a narrow compromise: it must be fast enough to reach the soliton step before the thermal shift moves the resonance, yet slow enough, and stopped at the right detuning, for the soliton to be captured and locked before thermal collapse occurs. Furthermore, even after successful initiation, maintaining the soliton state over extended durations requires continuous active feedback to counteract environmental perturbations and laser frequency drifts \cite{geng2020enhancing}.

 An initial step toward algorithmic automation was demonstrated using a genetic algorithm (GA) to optimize linear frequency tuning parameters (start \& stop frequencies and slew rate), providing flexibility in target comb states \cite{mazoukh2024genetic}. However, this approach remains restricted to basic linear frequency scans, and each realization takes approximately 12 hours to converge. This is dominated by the time required to repeatedly scan and poll an optical spectrum analyzer for every candidate solution.
Beyond frequency tuning, several alternative methods have been developed to bypass thermal instability by directly managing cavity temperature or intra-cavity power. For instance, auxiliary-laser-assisted schemes preserve thermal equilibrium by counter-propagating a secondary optical field \cite{zhou2019soliton}, but this requires an extra laser source. Similarly, active thermal tuning uses integrated micro-heaters to dynamically shift cavity resonance \cite{joshi2016thermally}. Automated soliton generation has recently been demonstrated via an active-kick technique using an FPGA to control these micro-heaters \cite{beals_kerr_2026}. However, this approach relies on strict heater current controls, and dedicated on-chip thermal infrastructure. Finally, power-kicking methods can be used for soliton triggering. They rely on ultra-fast power modulations to jump across thermal transients \cite{brasch_bringing_2016}.
However, this method requires an auxiliary modulator and precise control circuits that add hardware complexity and complicate integration. 
Even recent automated implementations that apply Q-learning on FPGAs remain fundamentally tied to power-kicking to initiate the soliton. They use RL solely to lengthen the step rather than addressing the underlying frequency-generation path \cite{wang2025fpga}. Alternatively, self-injection locking (SIL) passively locks the pump to a favorable cavity resonance via backscattered light, avoiding the need for continuous active feedback once locked. This mechanism, however, pins the pump frequency close to resonance, constraining both the accessible detuning range and comb bandwidth \cite{shen_integrated_2020, wang_integrated_2026}. Furthermore, SIL still requires coarse thermal tuning to pre-set the laser frequency close to a cavity resonance and turnkey soliton generation has a tendency to generate multi-solitons states \cite{shen_integrated_2020}. Consequently, triggering a single soliton requires careful tuning of the pump power, resonator temperature, or feedback optical phase \cite{shen_integrated_2020, wang_integrated_2026}.

These limitations motivate a control strategy that acts on the pump frequency alone and is not confined to a predefined scan. We propose RL for this purpose: an agent learns a control policy directly from interaction with the system, using measured outcomes as feedback and without requiring a model of the cavity dynamics \cite{sutton2018reinforcement}. Learning-based control has already been applied to nonlinear laser dynamics. Deep RL has been used to tune a spectral filter for stable dissipative soliton generation in a mode-locked fiber laser \cite{Kuprikov2022}, and to steer laser multistability toward a target harmonic mode-locked regime by jointly controlling pump power and saturable absorber transmission \cite{kokhanovskiy2024multistability}. However, deep architectures typically entail substantial computational overhead, a critical constraint for real-time physical control \cite{dulac-arnold_challenges_2021}. This motivates a lightweight agent that can learn online directly on embedded hardware.

We previously showed, in simulation only, that such an agent can navigate the bistability curve of a Lugiato-Lefever model to reach and maintain a target soliton spectrum \cite{sankar2025reinforcement}. Here, we implement it on a system-on-chip to control the pump detuning in real time. Because the agent adapts both the direction and magnitude of its tuning steps to the measured cavity state, it is not restricted to a linear sweep trajectory, and this additional degree of freedom allows it to outperform optimized linear and adaptive sweeps in our setup. Compared with the auxiliary-laser, micro-heater, and intensity-modulator schemes discussed above, our approach requires only control of the pump frequency. We demonstrate single-soliton generation in 4H-SiC microresonator with a 98\% success rate in an average of 1.32~s. Once captured, the soliton is maintained by a standard PID controller, and is autonomously reacquired after a forced collapse without manual intervention. These results establish RL-driven frequency tuning as a promising, hardware-efficient alternative for automated soliton generation in microring resonators.

\section{Experimental Setup}
\label{sec:experiment}

\begin{figure}[!htbp]
    \centering
    \includegraphics[width=\linewidth]{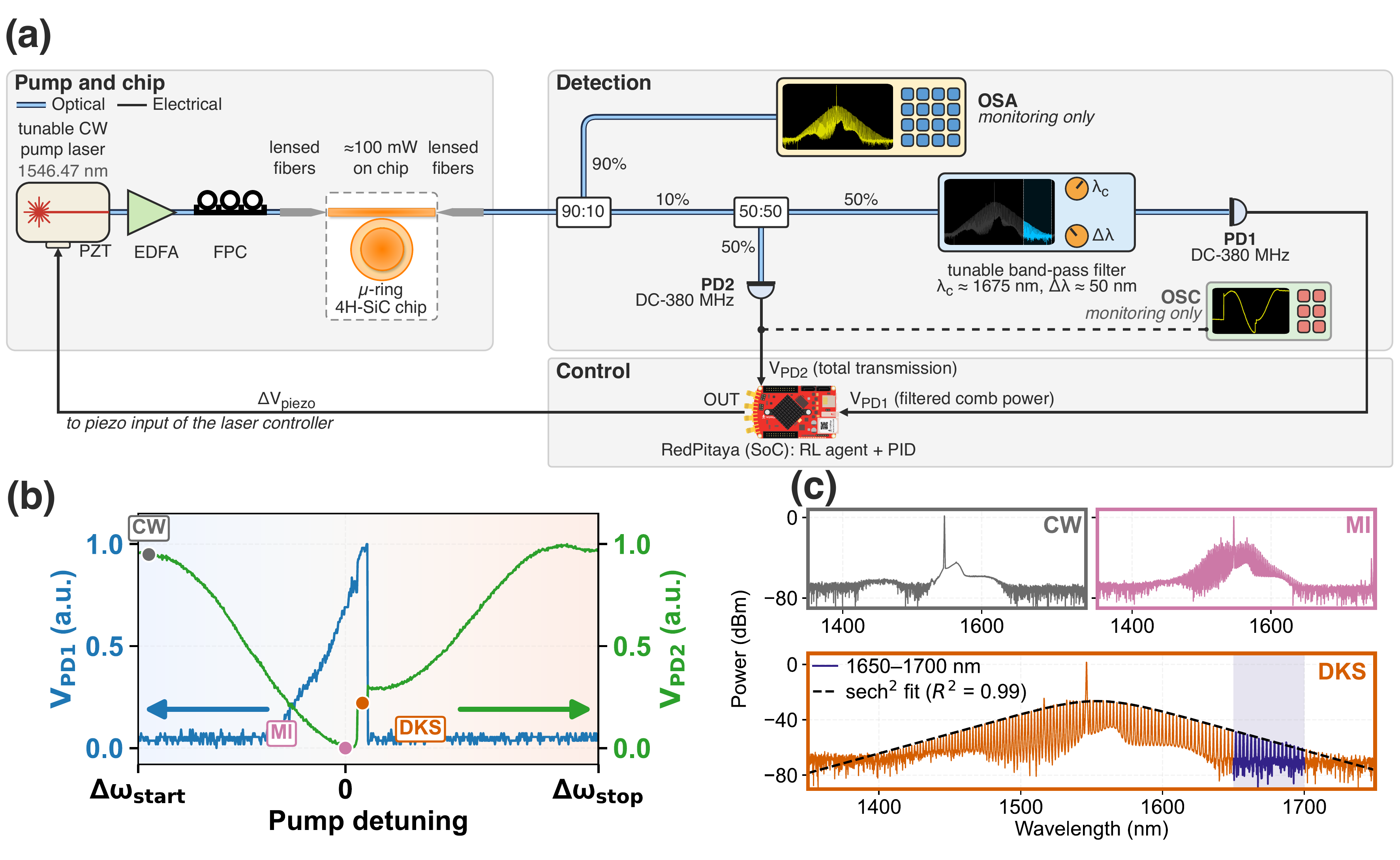}
\caption{Experimental setup and characteristic soliton signatures.
\textbf{(a)} Experimental setup.
\textbf{(b)} Cavity transmission ($\mathrm{V}_{\mathrm{PD2}}$) and filtered comb power ($\mathrm{V}_{\mathrm{PD1}}$) during a linear pump sweep, highlighting the different states (CW, MI and DKS).
\textbf{(c)} Corresponding optical spectra recorded on the OSA (RBW = 0.5~nm). The DKS spectrum is fitted with a $\text{sech}^2$ envelope (dashed line, $R^2 = 0.99$), confirming generation of a single dissipative Kerr soliton (DKS); the shaded band marks the spectral region (1650--1700~nm) selected by the band-pass filter for PD1.
Resonator ($\mathrm{TM}_{00}$, under-coupled): $R = 59.4~\mu\mathrm{m}$, FSR $\approx 294~\mathrm{GHz}$, $Q_\mathrm{L} = 1.94\times10^{6}$, $Q_\mathrm{i} = 3.12\times10^{6}$, $D_2/2\pi = 34~\mathrm{MHz}$. The device was fabricated with the process described in~\cite{Wang2026}.}
\label{fig:setup_and_spectrum}
\end{figure}

Our control framework uses only two photodetected signals: the total cavity transmission and the power in a narrow band on the long-wavelength side of the comb spectrum, selected by a single band-pass filter. Both are measured with photodiodes, so the feedback loop needs no optical spectrum analyzer, auxiliary laser, or high-speed detection. This simple, passive sensing stage is a natural candidate for on-chip integration. Fig.~\ref{fig:setup_and_spectrum} introduces the experimental platform used throughout this work: (a) the setup, (b) the characteristic signal traces used in our control approach, and (c) the corresponding optical spectra, including a typical single-soliton spectrum. The resonator used in our experiments is an integrated 4H-silicon-carbide (SiC) micro-ring. The main physical parameters are listed in the caption of Fig.~\ref{fig:setup_and_spectrum}, and fabrication details of this specific micro-ring have been described in Ref. \cite{Wang2026}. The optical pump path (Fig.~\ref{fig:setup_and_spectrum}(a)) originates from a tunable continuous-wave (CW) pump laser, amplified using an erbium-doped fiber amplifier (EDFA) and polarization-matched with a fiber polarization controller. The pump laser is tuned close to a resonance near 1546.47 nm and coupled into the chip via lensed fibers, with the on-chip pump power fixed at approximately 100 mW.

At the output of the micro-ring, the transmitted light is split into three paths. The first path is routed to an OSA for averaged recordings of the soliton spectrum (Fig.~\ref{fig:setup_and_spectrum}(c)) and is excluded from the feedback loop. The other two paths feed the feedback loop used for soliton generation and stabilization. One of them passes through a wavelength-tunable band-pass filter ($\lambda_c \approx 1675~\mathrm{nm}$, $\Delta\lambda_{\mathrm{FWHM}} \approx 50~\mathrm{nm}$) that selects a spectral band on the long-wavelength side of the comb, far from the pump (shaded band in Fig.~\ref{fig:setup_and_spectrum}(c)). This band is detected by $\mathrm{PD1}$ (DC to 380~MHz bandwidth), whose output voltage, $\mathrm{V}_{\mathrm{PD1}}$, is proportional to the optical power contained in the filtered band. The other path is sent directly to $\mathrm{PD2}$ (DC to 380~MHz bandwidth), whose output voltage, $\mathrm{V}_{\mathrm{PD2}}$, is proportional to the total transmitted optical power, i.e., the residual pump and the entire comb integrated over the PD bandwidth. Both voltages are digitized by a Red Pitaya STEMlab board\footnote{Red Pitaya STEMlab 125-14 PRO Gen 2 board, built around a Xilinx Zynq System-on-Chip (SoC). Analog input bandwidth DC to 60~MHz} running the RL agent and PID algorithms, and one output port of the board drives the laser's piezoelectric input to tune its frequency.

Linearly sweeping the pump detuning across the cavity resonance reveals distinct signatures in both feedback signals that uniquely identify each regime. As shown in Fig.~\ref{fig:setup_and_spectrum}(b), sweeping the laser detuning from blue to red causes the transmission signal ($\mathrm{V}_{\mathrm{PD2}}$) to drop, as optical power couples into the micro-ring. Simultaneously, the filtered power ($\mathrm{V}_{\mathrm{PD1}}$) rises due to spectral broadening during the modulation instability (MI) step, eventually reaching its maximum value at the level of the abrupt step in transmitted power corresponding to the DKS state, before dropping back when detuned out of resonance. The corresponding CW, MI and DKS spectra are shown in Fig.~\ref{fig:setup_and_spectrum}(c). In this resonator, such sweeps lead either to a single DKS or to collapse from the MI state to CW, with no multi-soliton steps observed. This is similar to the direct single-soliton access by slow pump tuning reported in Ref.\cite{sun2024directly}. Ultimately, these two signals serve as the primary observation space for the RL agent. They enable it to identify the cavity state and learn the optimal detuning commands required to reliably reach and maintain the soliton state.

\section{Reinforcement Learning Control Framework}
\label{sec:pump_detuning_agent}

\begin{figure}[!htbp]
    \centering
    \includegraphics[width=0.85\textwidth]{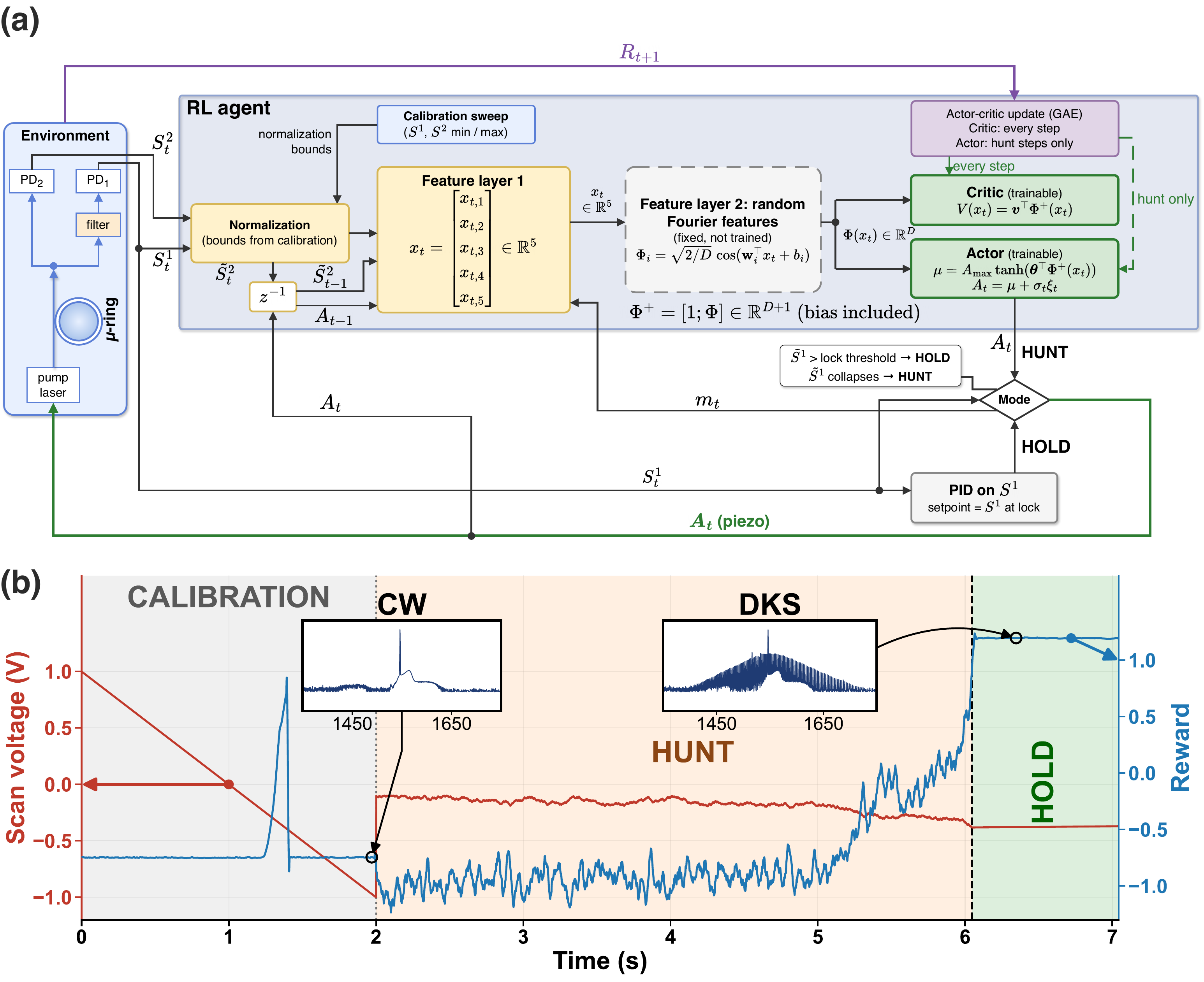}
    \caption{RL control framework. \textbf{(a)} Controller architecture on the Red Pitaya SoC. The measured signals $S_t^1$ and $S_t^2$ are normalized using bounds from the calibration sweep to give $\tilde S_t^1$ and $\tilde S_t^2$. Together with the previous voltage step and a hunt/hold flag, they form the state $\mathbf{x}_t\in\mathbb{R}^5$. Fixed random Fourier features map $\mathbf{x}_t$ to $\Phi(\mathbf{x}_t)\in\mathbb{R}^D$, and only the linear actor and critic weights are trained. In the hunt phase the actor outputs the piezo step $A_t=\Delta V_t$; once $\tilde S^1$ exceeds the lock threshold, a PID loop acting on $S^1_t$ takes over (hold) until the comb is lost. Optical and electrical paths are distinguished by line style. \textbf{(b)} Representative run showing the calibration, hunt and hold phases, with the scan voltage and reward traces and the CW and DKS spectra as insets.}
    \label{fig:software}
\end{figure}

The control problem is to determine how the pump laser frequency should be tuned across the cavity resonance so that the system is brought from the CW state into a single-soliton state and kept there, despite the thermal and non-linear dynamics described above. We formulate this as a reinforcement learning task, summarized in Fig.~\ref{fig:software}(a). The agent is the controller running on the Red Pitaya SoC, and the environment is the experimental system described in Sec.~\ref{sec:experiment}: the pump laser with its piezoelectric frequency tuning input, the microring resonator, and the two photodiodes that monitor the total transmission and the band-pass-filtered comb power. At each discrete time step $t$, the agent observes the measurement vector $\mathbf{S}_t=(S_t^1,S_t^2)$, where $S_t^1\equiv V_{\mathrm{PD1}}(t)$ is the filtered comb power and $S_t^2\equiv V_{\mathrm{PD2}}(t)$ is the cavity transmission, i.e., the digitized photodiode voltages of Sec.~\ref{sec:experiment} sampled at step $t$. From these measurements the agent constructs a \emph{state} $\mathbf{x}_t$ and decides on an \emph{action} $A_t$. The environment then responds with a new observation $\mathbf{S}_{t+1}$ and a scalar \emph{reward} $R_{t+1}$, a number that scores how much the action moved the system toward the single-soliton state. This closes a loop between the microring dynamics and the agent. The agent's goal is to learn the \emph{policy}, i.e., the mapping from states to tuning actions, that maximizes the reward accumulated over time.

The action $A_t=\Delta V_t$ is the instantaneous tuning voltage step applied to the pump laser's piezoelectric actuator. The signals ($\mathrm{V}_{\mathrm{PD1}} ~ \& ~\mathrm{V}_{\mathrm{PD2}}$) are first min--max normalized to $[-1,1]$ using bounds obtained from an initial calibration sweep (Fig.~\ref{fig:software}(a), top, and the grey region in Fig.~\ref{fig:software}(b)), giving $\tilde S_t^{i}$ for $i=1,2$. The state $\mathbf{x}_t\in\mathbb{R}^5$ is a compact, control-oriented representation of the cavity. It contains the normalized total transmission $\tilde S_t^2$, its step-to-step difference $\tilde S_t^2-\tilde S_{t-1}^2$, and the filtered comb power $\tanh(\tilde S_t^1)$, where the $\tanh$ softly bounds readings that exceed the calibrated range. It also contains the voltage step applied at the previous step, $A_{t-1}$, to retain memory of the tuning trajectory, and a binary flag $m_t$ indicating hunt or hold mode (see supplemental Section~S1.1). Because the filtered comb power is maximal in the soliton state (Fig.~\ref{fig:setup_and_spectrum}(b)), it also serves as the primary reward signal (Eq.~\eqref{eq:reward}) to encourage soliton acquisition and stabilization.

To map $\mathbf{x}_t$ to tuning actions, we employ a lightweight Actor-Critic architecture \cite{konda1999actor} (detailed in Section~S1.2), chosen to keep training sample-efficient enough for direct deployment on the SoC. As shown in Fig.~\ref{fig:software}(a), the state is first projected onto a $D$-dimensional nonlinear feature space using fixed random Fourier features \cite{rahimi2007random}, $\Phi(\mathbf{x}_t)=\sqrt{2/D}\,\cos(\mathbf{W}\mathbf{x}_t+\mathbf{b})$, with $D=16$. Here $\mathbf{W}\in\mathbb{R}^{D\times5}$ is a random projection matrix and $\mathbf{b}\in\mathbb{R}^{D}$ a random phase offset, both drawn once at initialization and never trained (Section~S1.1, Eq.~(S3)). A linear actor and a linear critic then operate on $\Phi^{+}=[1;\Phi]\in\mathbb{R}^{D+1}$, where the constant entry provides a bias term. The actor outputs the mean $\mu(\mathbf{x}_t)=A_{\max}\tanh\!\big(\boldsymbol{\theta}^{\top}\Phi^{+}(\mathbf{x}_t)\big)$ of a Gaussian policy over the voltage step, to which an exploration noise that decays during the hunt is added. The executed step is clipped to a maximum amplitude $A_{\max}$. The critic does not act on the cavity. It estimates the expected return $V(\mathbf{x}_t)=\mathbf{v}^{\top}\Phi^{+}(\mathbf{x}_t)$ from the current state, and the difference between the received reward and this estimate (the advantage, computed with generalized advantage estimation \cite{schulman2015high}) indicates whether a step was better or worse than expected. This advantage function is used to update the actor, which reinforces steps that outperform the critic's expectation, while the critic itself is trained to predict the return more accurately. The critic is therefore only needed during learning and does not enter the control action. The weight vectors $\boldsymbol{\theta},\mathbf{v}\in\mathbb{R}^{D+1}$ are the only trained parameters, i.e., $2(D+1)=34$ in total. Such lightness is essential because the agent learns entirely online from live hardware interaction, where every trial consumes real experimental time rather than cheap simulated steps.


This same constraint shapes how the reward itself is designed, combining the filtered comb power $S^1_t$ with two shaping terms that encode prior physical knowledge to accelerate convergence. It is expressed as
\begin{equation}
    R_{t+1}=\tilde{P}_{\text{comb}}+\mathcal{G}_{\text{trans}}+\mathcal{D}_{\text{dir}},
    \label{eq:reward}
\end{equation}
where $\tilde{P}_{\text{comb}}=\tanh(\tilde S_{t+1}^1)$ is the filtered comb power, rescaled using the power boundaries identified during calibration. Since the calibration curve (Fig.~\ref{fig:setup_and_spectrum}(b)) shows that the soliton step is reached by tuning from the blue to the red detuned side of the resonance, $\mathcal{G}_{\text{trans}}$ penalizes dwelling in the high-transmission, blue-detuned region, while $\mathcal{D}_{\text{dir}}$ applies a directional bias toward red detuning based on the sign of the applied action. These terms encode only the known direction of the soliton step, not a trajectory. Their weight is chosen to be large enough to bias exploration toward the correct side of the resonance, yet small enough (about 20\% of the total reward, Supplement S1.2) that the comb-power term dominates and the agent can deviate from this prior when it is beneficial. This is a standard use of known system structure to reduce sample complexity in reinforcement learning, particularly valuable given the constrained real-time training budget on embedded hardware \cite{ng1999policy}. Rather than relying on pre-calculated transition maps or rigid heuristics, the fine-tuning dynamics are learned continuously through direct interaction with the cavity.

The state-action-reward loop is executed on the SoC in three phases, illustrated by a representative run in Fig.~\ref{fig:software}(b). The SoC output is limited to $\pm1$~V, so the laser controller (Toptica DLC) is configured with an offset of 70~V and a gain of 70~V/V, which maps this range onto the full 0--140~V span of the piezoelectric actuator. The scan voltage in Fig.~\ref{fig:software}(b) is expressed in SoC output units.

Operation begins with a linear \textbf{calibration} sweep over the full output range (as in Fig.~\ref{fig:software}(b)), which serves two purposes. First, it establishes the normalization bounds ($S^i_{\min}$, $S^i_{\max}$) of the two photodetected signals. Second, it locates the soliton step, i.e., the piezo voltage at which $S^1_t$ is maximal, and uses it to restrict the search range of the subsequent hunt to a narrow window around this step, covering about one third of the calibration range. Finding where the soliton step lies is therefore straightforward and requires no learning. The remaining difficulty is dynamical. The actuator cannot be slewed arbitrarily fast without exciting its mechanical resonances, while a slow approach gives the thermal shift time to push the system out of the soliton existence range after the abrupt power drop. The agent must therefore learn a detuning trajectory that balances approach speed against these two constraints.

The system then restarts from the continuous-wave (CW) state, in which the pump is coupled into the resonator without comb generation, at the blue-detuned edge of the window, close to but not at the soliton step, and enters the \textbf{hunt} phase ($m_t=-1$). Here the RL agent acts on the pump detuning over successive steps until a soliton is found, defined as $\tilde S^1$ exceeding 90\% of its calibrated range. This transition to the DKS state is characterized by a sharp rise of the reward signal, at which point the Mode switch in Fig.~\ref{fig:software}(a) hands control to the \textbf{hold} phase ($m_t=+1$). A PID controller then regulates the pump detuning using the filtered comb power $S^1_t$ as the process variable, with the setpoint taken at lock, while the actor is idle. Should the comb subsequently be lost, $S^1_t$ collapses to the bottom of its calibrated range rather than merely dipping below the acquisition threshold. This collapse triggers the return to the hunt phase, allowing the agent to reacquire the soliton. Because the locked state is sustained at high filtered power and the loss is a near-total collapse, the trigger is unambiguous and is not falsely activated by minor fluctuations of $S^1_t$ near the acquisition threshold.

This architecture therefore decouples the two competing demands of the problem. The hunt phase handles the nonlinear, transient dynamics of soliton formation, while the hold phase entrusts long-term retention to a simple, well-understood linear controller once the system is already close to its target operating point.

\section{Results}

\subsection{Comparison with Open-Loop Baselines}
\label{subsec:macroscopic_reliability}
Having described the operation of the framework over a single representative run, we now assess its performance statistically across many repeated trials. We first benchmark the agent's overall success rate against fixed-slope linear frequency sweeps and an additive increase/multiplicative decrease (AIMD) linear scan, whose slope is tuned from trial to trial. We then quantify the average time required to capture the soliton state, and finally examine the impact of the PID (hold phase) on the comb retention time once it has been generated.

To evaluate the framework's overall reliability, the RL agent was tested over two consecutive campaigns of 50 trials each (run 1 and run 2), separated by a gap of only a few minutes, with no parameter intervention or manual recalibration between or within campaigns, for a total of 100 trials. Two classes of baseline were implemented for comparison, both using a two-phase architecture identical to the RL setup in which the active hunt phase is replaced by a linear detuning scan while the hold phase remains unchanged. The fixed-slope scans were each evaluated over 50 trials, whereas the AIMD linear scan described below was evaluated under the same two-campaign, 50-trial protocol as the RL agent (100 trials), with its success rate taken over both campaigns.
Both baselines share the same lock criterion as the RL agent: if the filtered comb power $S^1_t$ exceeds 90\% of its calibrated range ($\tilde S^1 \ge 0.8$), the algorithm assumes a soliton step has been found and engages the PID lock. A trial is considered successful if the comb is generated and the PID lock sustains it for 10~s, about ten times the median RL capture time reported below. This ensures that the soliton survives well beyond the capture dynamics and the thermal transient that follows the abrupt power drop, while keeping each trial short enough to collect many trials per method.
The first approach uses fixed-slope scans at three representative speeds manually selected, following typical experimental protocols, to be fast enough for soliton generation while remaining within the piezo's optimal operating limit. The second is a scan governed by an AIMD rule \cite{chiu1989analysis,jacobson1988congestion}, in which the step size of the linear ramp is adapted trial-to-trial rather than fixed by hand. We adopt AIMD because it is a standard, model-free rule for probing the maximum tolerable rate of a resource from binary success/failure feedback, as in network congestion control. For fair comparison, the AIMD step size is upper-bounded by the same maximum per-step size that constrains the RL agent. More precisely, for a successful trial, the step size is increased by a small additive amount. On the contrary, when unsuccessful, whether the sweep missed the soliton step entirely or the comb was found but subsequently dropped during the hold, the step size is instead halved (see supplemental Section~S2). In parallel, the scan window is re-centered after each successful trial on the detuning voltage at which the comb was last triggered, compensating for thermal drift over the course of the run. Both the RL agent and the AIMD baseline continue to update their parameters online throughout each evaluation campaign, so the reported success rates reflect performance under continual adaptation rather than a frozen, pre-trained policy. 

\begin{figure}[!htb]
\centering
\includegraphics[width=\linewidth]{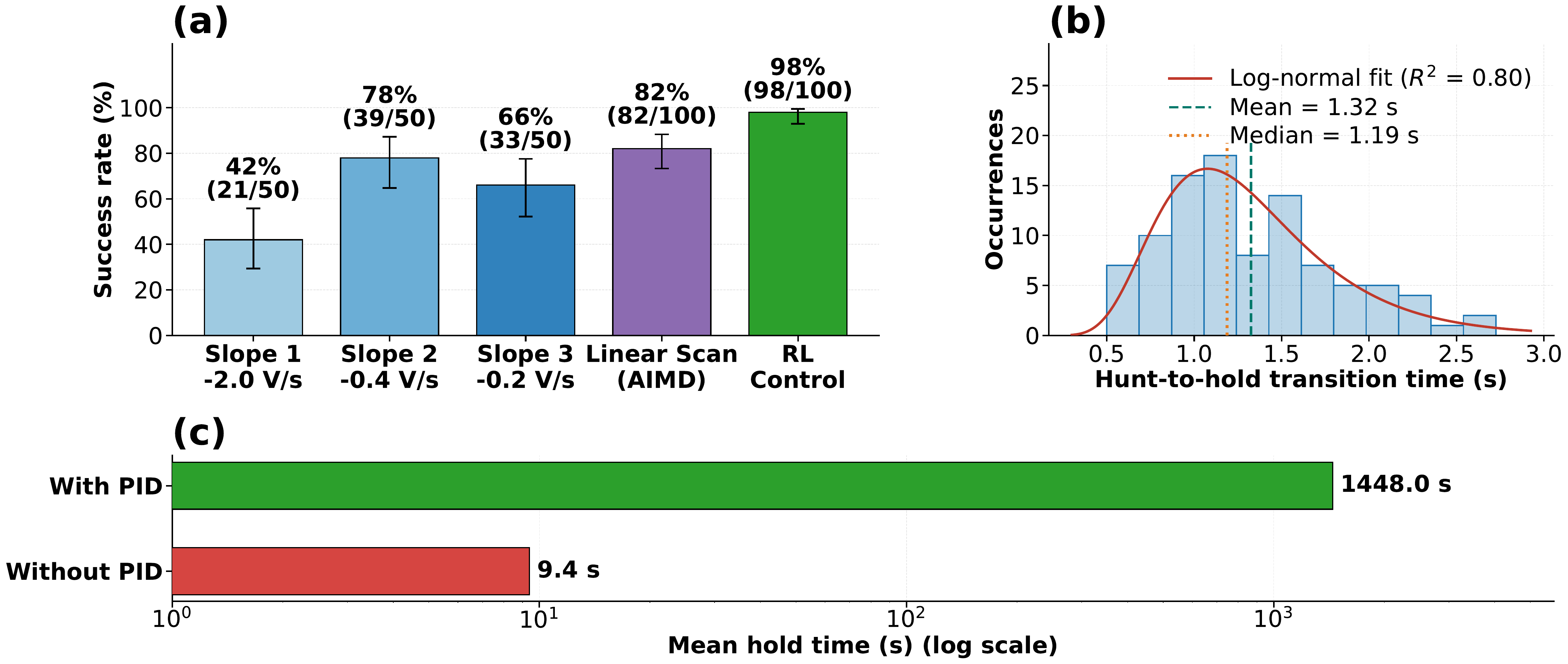}
\caption{
    Performance evaluation of the RL-based comb generation and PID stabilization.
    \textbf{(a)} Comb-lock success rates comparing fixed-slope linear sweeps ($-0.2$, $-0.4$, and $-2.0\text{ V/s}$), a tuned AIMD linear scan, and the RL control agent.
    \textbf{(b)} Distribution and log-normal fit of the hunt-to-hold transition times for successful RL trials, indicating the mean ($\mu = 1.32\text{ s}$) and median ($1.19\text{ s}$).
    \textbf{(c)} Comparison of average comb hold duration with and without active PID stabilization.
}
\label{fig:trajectories_and_stats}
\end{figure}

Fig.~\ref{fig:trajectories_and_stats}(a) summarizes the aggregated success probabilities for these campaigns, with 95\% Wilson score intervals included to account for binomial proportion uncertainty near the boundaries (see supplemental Section~S3). Since the RL agent actively modulates its tuning trajectory, it possesses no fixed slope, so we can only compute an average sweep rate ($-0.22\,\text{V/s}$). The fixed-slope linear scans were evaluated across three speeds, comparable to the RL average (slope 3, $-0.2\,\text{V/s}$), faster (slope 2, $-0.4\,\text{V/s}$), and an order of magnitude faster (slope 1, $-2.0\,\text{V/s}$), each run over 50 independent trials, for a total of 150 fixed-slope baseline trials.

The RL agent achieves a robust 98.0\% overall success rate, consistent across both back-to-back 50-trial campaigns (runs 1 and 2), indicating stable performance rather than a result specific to a single run. The fixed-slope linear scans max out at 78.0\% even under the best hand-tuned speed, while the AIMD scan improves on this, reaching 82.0\%, confirming that letting the scan speed adapt to trial outcomes is preferable to any single hand-picked rate. The 95\% Wilson score intervals for the RL agent (98.0\%, [93.0\%, 99.4\%]) and the AIMD baseline (82.0\%, [73.3\%, 88.3\%]) do not overlap, indicating that this improvement is statistically robust rather than attributable to sampling variation across the two 50-trial campaigns. However, the AIMD scan is limited to a straight, monotonic ramp within any given trial, since only its step size and window position evolve, and only from one trial to the next, whereas the RL agent computes a distinct non-linear trajectory tailored to the measured cavity state on every attempt. The higher success rate therefore indicates that the benefit lies in shaping the trajectory within a single attempt, rather than in the choice of an average speed: the agent can vary its step size as the cavity approaches the soliton step, which a constant-rate ramp cannot do. Fig.~\ref{fig:trajectories_and_stats}(b) details the time taken for the active RL hunt phase across all successful trials, yielding an average capture time of 1.32 seconds. This timescale is set by the mandatory delays between steps, necessary to avoid mechanical resonance in the piezoelectric actuator tuning the pump frequency, rather than by the computational overhead of the digital control loop, which completes each policy evaluation and update well within the available timing budget. A detailed breakdown of failure mechanisms for each method is provided in supplemental Section~S2.

To demonstrate the critical role of the secondary PID lock phase in combating environmental noise, the comb lifetime was evaluated with and without active stabilization. In both cases, once a soliton was generated via the RL agent, the system was left completely unperturbed until the state naturally decayed. As shown in Fig.~\ref{fig:trajectories_and_stats}(c), engaging the PID controller dramatically extends the operational lifespan, increasing the average comb hold time by a factor of $\approx 154$. The eventual loss of lock in the PID-stabilized case is attributed to slow coupling misalignment and polarization drift rather than to any limitation of the PID loop itself.

\subsection{Robustness to Physical Disturbances}
\label{subsec:robustness}
\begin{figure}[!htbp]
    \centering
    \includegraphics[width=\linewidth]{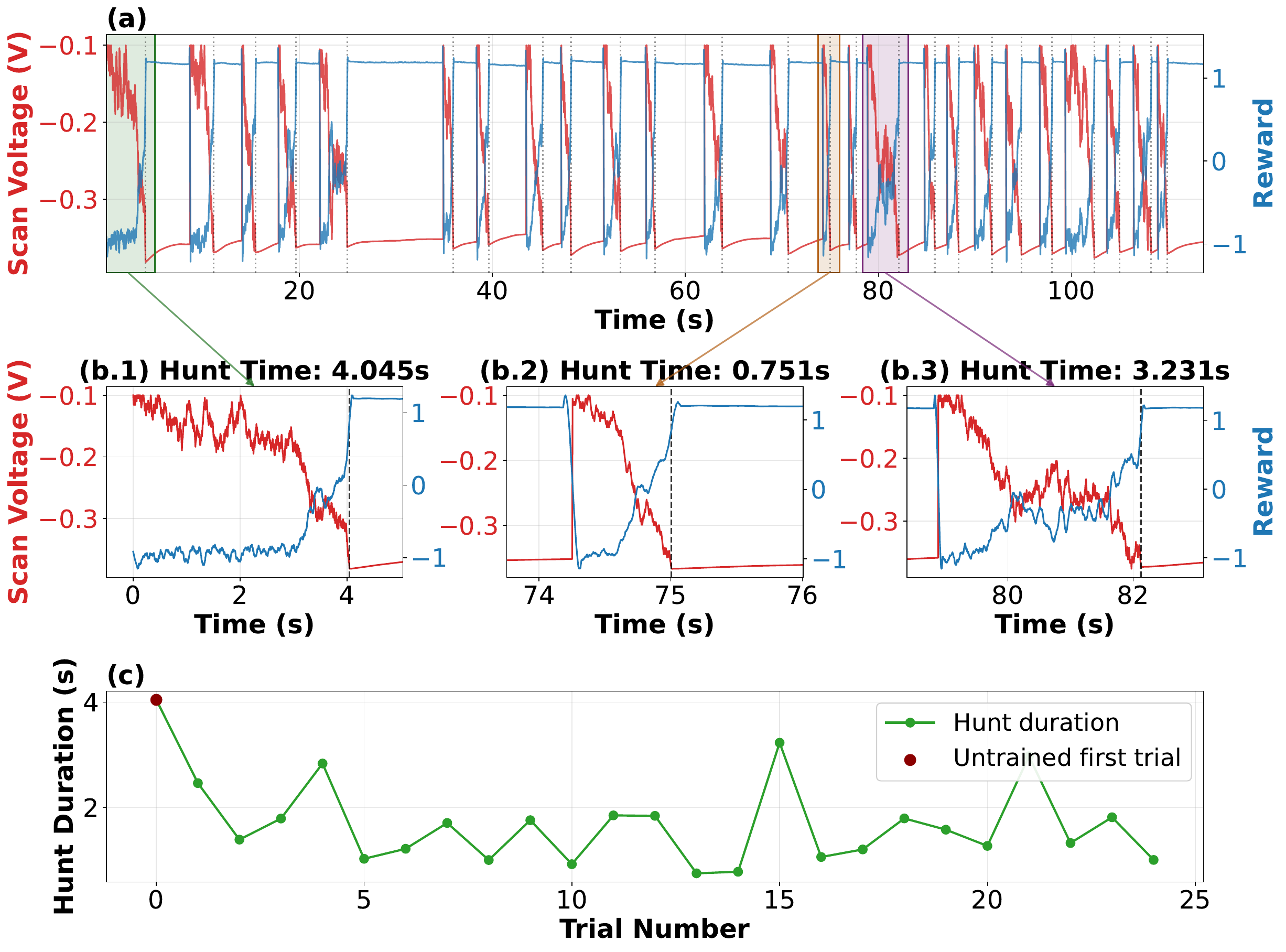}
    \caption{
        Robustness of the RL control framework against intentional (uncalibrated) physical disturbances. \textbf{(a)} Macroscopic time-series of the scan voltage and reward signal, showing multiple automated recoveries after the soliton state is forcefully disrupted. The dashed vertical lines indicate the hunt-to-hold transition points. \textbf{(b)} Representative hunt-phase transitions: \textbf{(b.1)} the initial, untrained trial; \textbf{(b.2)} the shortest hunt duration among subsequent recapture events; \textbf{(b.3)} the longest hunt duration among subsequent recapture events, demonstrating the range of recapture speeds the agent achieves once trained. \textbf{(c)} Hunt duration across all recapture trials, remaining consistently well below the untrained first trial throughout continued operation.
    }
    \label{fig:disturbance_recovery}
\end{figure}

Having established the framework's performance under nominal operating conditions, we evaluate the real-time performance and adaptive learning capabilities of the deployed RL agent by examining its ability to autonomously recover the DKS state under continuous operation and intentional physical disturbance (tapping on the optical table). Fig.~\ref{fig:disturbance_recovery}(a) shows the agent's action and the reward signal when the soliton state is forcefully disrupted during a continuous operation.

It is observed in Fig.~\ref{fig:disturbance_recovery}(b.1) and (c) that the very first hunt phase, beginning from a completely untrained agent state, exhibits the longest \textbf{hunt} duration of the entire trial sequence ($4\text{ s}$). This extended duration was observed in all experiments when the algorithm is initialized. This is expected, as the agent has no prior experience and must rely on initial active exploration to discover the optimal tuning trajectory. Following this initial capture, mechanical perturbations are sequentially introduced by tapping on the optical table. The shocks physically disrupt the cavity stability and induce a sudden dropout from the locked state, causing an immediate collapse in the comb power.

Rather than requiring a manual reset, this loss event is immediately detected by the control framework, which automatically returns the system to the hunt phase. Once trained, the agent converges toward faster recapture behavior, with recapture durations remaining well below the untrained first trial throughout the remainder of the trial sequence (Fig.~\ref{fig:disturbance_recovery}(c)). Among these recapture events, the longest hunt duration is $3.231\text{ s}$ (Fig.~\ref{fig:disturbance_recovery}(b.3)), still shorter than the untrained first trial, while the shortest is $0.751\text{ s}$ (Fig.~\ref{fig:disturbance_recovery}(b.2)), a more than five times reduction relative to the first trial, with occasional excursions corresponding to more severe disturbances.

The agent's detuning trajectories vary substantially from one recapture event to the next, rather than converging on a single repeated pattern. This is evident in Fig.~\ref{fig:disturbance_recovery}(b.1--b.3), where the scan voltage traces differ markedly in shape, speed, and duration despite all successfully locating the same soliton step, and further confirmed by the absence of a fixed slope in the aggregated trajectory in Fig.~\ref{fig:disturbance_recovery}(a). This variability shows that the agent is not executing a conventional, fixed-rate linear scan learned once and repeated. Instead, it computes a distinct, non-linear trajectory for each capture attempt, tailored to the cavity state observed.

These results indicate that the agent's freedom to depart from a fixed detuning trajectory contributes to its higher success rate relative to every fixed-slope scan and to its ability to reacquire the soliton after disturbances. The OSA spectrum during repeated generation and recovery of the soliton is shown in \href{https://opticapublishing.figshare.com/s/7bda7eee70c02b36a338?file=71903641}{Visualization 1} (supplemental material).

\section{Conclusion}
We have shown that a lightweight reinforcement learning agent combined with a PID controller can generate and retain a single soliton in a 4H-SiC microring using only pump-frequency tuning, one optical band-pass filter and two photodiodes. The agent captures the soliton in 1.32~s on average with a 98\% success rate, compared with up to 78\% for fixed-slope linear sweeps and 82\% for AIMD scan, an advantage we attribute to its ability to adapt the detuning trajectory within each capture attempt. Once captured, the soliton is retained by the PID loop for 1448~s on average, a $\approx$154-fold improvement over the unstabilized case, and it is autonomously reacquired after forced collapses without manual intervention. Because the actor and critic are linear readouts over a fixed random feature map (34 trained parameters), the policy learns online and runs entirely on an embedded SoC, avoiding the GPU and offline training typical of deep RL in related laser work \cite{Kuprikov2022,kokhanovskiy2024multistability}. The 1.32~s capture time is set by the mechanical response of the piezoelectric actuator rather than by the algorithm or the SoC, leaving substantial headroom for faster capture with a faster actuator. Together with a passive, low-bandwidth sensing stage, this offers a plausible path toward co-integration and portable microcomb sources.

Future work includes extending the reward formulation toward multi-soliton and soliton-crystal states, mirroring how the GA approach can be redirected via its fitness function, and validating the framework across different resonators to establish applicability beyond the platform studied here.

\begin{backmatter}

\bmsection{Supplementary Information}
See Supplement 1 for supporting content.

\bmsection{Funding}
European Union's Horizon Europe Research and Innovation Programme, Marie Skłodowska-Curie Grant Agreement No.~101169370 and Marie Skłodowska-Curie Grant Agreement No.~101227010. Villum Foundation, VI-POPCOM Project (VIL54486). European Union’s Horizon 2020 FET Open (SiComb, No. 899679),


\bmsection{Disclosures}
The authors declare no conflicts of interest.

\bmsection{Data Availability}
Data underlying the results presented in this paper are not publicly available at this time but may be obtained from the authors upon reasonable request.

\end{backmatter}



\bibliography{sample}

\begin{thebibliography}{10}
\newcommand{\enquote}[1]{``#1''}

\bibitem{diddams2020optical}
S.~A. Diddams, K.~Vahala, and T.~Udem, \enquote{Optical frequency combs: Coherently uniting the electromagnetic spectrum,} {\protect\JournalTitle{Science}} \textbf{369}, eaay3676 (2020). {\href{https://doi.org/10.1126/science.aay3676}{doi:10.1126/science.aay3676}}.

\bibitem{pasquazi2018microcombs}
A.~Pasquazi, M.~Peccianti, L.~Razzari, \emph{et~al.}, \enquote{Micro-combs: A novel generation of optical sources,} {\protect\JournalTitle{Physics Reports}} \textbf{729}, 1--60 (2018). {\href{https://doi.org/10.1016/j.physrep.2017.08.004}{doi:10.1016/j.physrep.2017.08.004}}.

\bibitem{herr2013si}
T.~Herr, V.~Brasch, J.~D. Jost, \emph{et~al.}, \enquote{Temporal solitons in optical microresonators,} {\protect\JournalTitle{Nature Photonics}} \textbf{8}, 145--152 (2014). {\href{https://doi.org/10.1038/nphoton.2013.343}{doi:10.1038/nphoton.2013.343}}.

\bibitem{sun2023review}
Y.~Sun, J.~Wu, M.~Tan, \emph{et~al.}, \enquote{Applications of optical microcombs,} {\protect\JournalTitle{Advances in Optics and Photonics}} \textbf{15}, 86--175 (2023). {\href{https://doi.org/10.1364/AOP.470264}{doi:10.1364/AOP.470264}}.

\bibitem{corcoran2025optical}
B.~Corcoran, A.~Mitchell, R.~Morandotti, \emph{et~al.}, \enquote{Optical microcombs for ultrahigh-bandwidth communications,} {\protect\JournalTitle{Nature Photonics}} \textbf{19}, 451--462 (2025). {\href{https://doi.org/10.1038/s41566-025-01662-9}{doi:10.1038/s41566-025-01662-9}}.

\bibitem{carmon2004dynamical}
T.~Carmon, L.~Yang, and K.~J. Vahala, \enquote{Dynamical thermal behavior and thermal self-stability of microcavities,} {\protect\JournalTitle{Optics Express}} \textbf{12}, 4742--4750 (2004). {\href{https://doi.org/10.1364/OPEX.12.004742}{doi:10.1364/OPEX.12.004742}}.

\bibitem{bao2017direct}
C.~Bao, Y.~Xuan, J.~A. Jaramillo-Villegas, \emph{et~al.}, \enquote{Direct soliton generation in microresonators,} {\protect\JournalTitle{Optics Letters}} \textbf{42}, 2519--2522 (2017). {\href{https://doi.org/10.1364/OL.42.002519}{doi:10.1364/OL.42.002519}}.

\bibitem{yi2016active}
X.~Yi, Q.-F. Yang, K.~Y. Yang, and K.~Vahala, \enquote{Active capture and stabilization of temporal solitons in microresonators,} {\protect\JournalTitle{Optics Letters}} \textbf{41}, 2037--2040 (2016). {\href{https://doi.org/10.1364/OL.41.002037}{doi:10.1364/OL.41.002037}}.

\bibitem{geng2020enhancing}
Y.~Geng, W.~Cui, J.~Sun, \emph{et~al.}, \enquote{Enhancing the long-term stability of dissipative kerr soliton microcomb,} {\protect\JournalTitle{Optics Letters}} \textbf{45}, 5073--5076 (2020). {\href{https://doi.org/10.1364/OL.400656}{doi:10.1364/OL.400656}}.

\bibitem{mazoukh2024genetic}
C.~Mazoukh, L.~Di~Lauro, I.~Alamgir, \emph{et~al.}, \enquote{Genetic algorithm-enhanced microcomb state generation,} {\protect\JournalTitle{Communications Physics}} \textbf{7}, 81 (2024). \href{https://doi.org/10.1038/s42005-024-01558-0}{doi:10.1038/s42005-024-01558-0}.

\bibitem{zhou2019soliton}
H.~Zhou, Y.~Geng, W.~Cui, \emph{et~al.}, \enquote{Soliton bursts and deterministic dissipative kerr soliton generation in auxiliary-assisted microcavities,} {\protect\JournalTitle{Light: Science \& Applications}} \textbf{8}, 50 (2019). {\href{https://doi.org/10.1038/s41377-019-0161-y}{doi:10.1038/s41377-019-0161-y}}.

\bibitem{joshi2016thermally}
C.~Joshi, J.~K. Jang, K.~Luke, \emph{et~al.}, \enquote{Thermally controlled comb generation and soliton modelocking in microresonators,} {\protect\JournalTitle{Optics Letters}} \textbf{41}, 2565--2568 (2016). {\href{https://doi.org/10.1364/OL.41.002565}{doi:10.1364/OL.41.002565}}.

\bibitem{beals_kerr_2026}
G.~J. Beals, Y.~Zhao, S.~K. Dacha, \emph{et~al.}, \enquote{Kerr soliton generation in ultra-compact photonic devices,} arXiv:2606.18067 [physics.optics] (2026). {\href{https://doi.org/10.48550/arXiv.2606.18067}{doi:10.48550/arXiv.2606.18067}}.

\bibitem{brasch_bringing_2016}
V.~Brasch, M.~Geiselmann, M.~H.~P. Pfeiffer, and T.~J. Kippenberg, \enquote{Bringing short-lived dissipative {Kerr} soliton states in microresonators into a steady state,} {\protect\JournalTitle{Optics Express}} \textbf{24}, 29312--29320 (2016). {\href{https://doi.org/10.1364/OE.24.029312}{doi:10.1364/OE.24.029312}}.

\bibitem{wang2025fpga}
Z.~Wang, J.~Baek, C.~Ahn, \emph{et~al.}, \enquote{Fpga-driven soliton locking for reduced manual intervention in microcombs,} {\protect\JournalTitle{Journal of Lightwave Technology}}  (2025). {\href{https://doi.org/10.1109/JLT.2025.3600657}{doi:10.1109/JLT.2025.3600657}}.

\bibitem{shen_integrated_2020}
B.~Shen, L.~Chang, J.~Liu, \emph{et~al.}, \enquote{Integrated turnkey soliton microcombs,} {\protect\JournalTitle{Nature}} \textbf{582}, 365--369 (2020). {\href{https://doi.org/10.1038/s41586-020-2358-x}{doi:10.1038/s41586-020-2358-x}}.

\bibitem{wang_integrated_2026}
Z.~Wang, T.~Xu, Y.~Wang, \emph{et~al.}, \enquote{Integrated soliton microcombs beyond the turnkey limit,} {\protect\JournalTitle{Physical Review Letters}} \textbf{137}, 043802 (2026). {\href{https://doi.org/10.1103/r8fh-vtps}{doi:10.1103/r8fh-vtps}}.

\bibitem{sutton2018reinforcement}
R.~S. Sutton and A.~G. Barto, \emph{Reinforcement Learning: An Introduction} (MIT Press, Cambridge, MA, 2018), 2nd ed. {\url{https://mitpress.mit.edu/9780262039246/reinforcement-learning/}}.

\bibitem{Kuprikov2022}
E.~Kuprikov, A.~Kokhanovskiy, K.~Serebrennikov, and S.~K. Turitsyn, \enquote{Deep reinforcement learning for self-tuning laser source of dissipative solitons,} {\protect\JournalTitle{Scientific Reports}} \textbf{12}, 7185 (2022). {\href{https://doi.org/10.1038/s41598-022-11274-w}{doi:10.1038/s41598-022-11274-w}}.

\bibitem{kokhanovskiy2024multistability}
A.~Kokhanovskiy, E.~Kuprikov, K.~Serebrennikov, \emph{et~al.}, \enquote{Multistability manipulation by reinforcement learning algorithm inside mode-locked fiber laser,} {\protect\JournalTitle{Nanophotonics}} \textbf{13}, 2891--2901 (2024). {\href{https://doi.org/10.1515/nanoph-2023-0792}{doi:10.1515/nanoph-2023-0792}}.

\bibitem{dulac-arnold_challenges_2021}
G.~Dulac-Arnold, N.~Levine, D.~J. Mankowitz, \emph{et~al.}, \enquote{Challenges of real-world reinforcement learning: definitions, benchmarks and analysis,} {\protect\JournalTitle{Machine Learning}} \textbf{110}, 2419--2468 (2021). {\href{https://doi.org/10.1007/s10994-021-05961-4}{doi:10.1007/s10994-021-05961-4}}.

\bibitem{sankar2025reinforcement}
V.~Sankar, C.~Peucheret, and D.~Zibar, \enquote{Reinforcement learning for soliton generation in micro-ring resonators,} in \emph{2025 IEEE Photonics Conference (IPC),}  (2025), p. WH4.5. {\href{https://doi.org/10.1109/IPC65510.2025.11282300}{doi:10.1109/IPC65510.2025.11282300}}.

\bibitem{Wang2026}
Y.~Wang, S.~Zhang, Y.~Lu, \emph{et~al.}, \enquote{Thermal behavior analysis of raman lasing in 4h-sic microresonators,} {\protect\JournalTitle{Applied Physics Letters}} \textbf{128} (2026). {\href{https://doi.org/10.1063/5.0318996}{doi:10.1063/5.0318996}}.

\bibitem{sun2024directly}
W.~Sun, J.~Li, R.~Wang, and Q.~Li, \enquote{Directly accessing the single-soliton state of a kerr microcomb and its universal scaling law,} {\protect\JournalTitle{Optical Materials Express}} \textbf{14}, 2938--2948 (2024). {\href{https://doi.org/10.1364/OME.541224}{doi:10.1364/OME.541224}}.

\bibitem{konda1999actor}
V.~R. Konda and J.~N. Tsitsiklis, \enquote{Actor-critic algorithms,} in \emph{Advances in Neural Information Processing Systems,}  vol.~12 (1999), pp. 1008--1014. {\href{https://dl.acm.org/doi/10.5555/3009657.3009799}{doi:10.5555/3009657.3009799}}.

\bibitem{rahimi2007random}
A.~Rahimi and B.~Recht, \enquote{Random features for large-scale kernel machines,} in \emph{Advances in Neural Information Processing Systems,}  vol.~20 (2007), pp. 1177--1184. {\href{https://dl.acm.org/doi/10.5555/2981562.2981710}{doi:10.5555/2981562.2981710}}.

\bibitem{schulman2015high}
J.~Schulman, P.~Moritz, S.~Levine, \emph{et~al.}, \enquote{High-dimensional continuous control using generalized advantage estimation,} arXiv:1506.02438 (2015). {\url{https://arxiv.org/abs/1506.02438}}.

\bibitem{ng1999policy}
A.~Y. Ng, D.~Harada, and S.~J. Russell, \enquote{Policy invariance under reward transformations: Theory and application to reward shaping,} in \emph{Proceedings of the Sixteenth International Conference on Machine Learning,}  (Morgan Kaufmann Publishers Inc., San Francisco, CA, USA, 1999), ICML '99, p. 278–287.

\bibitem{chiu1989analysis}
D.-M. Chiu and R.~Jain, \enquote{Analysis of the increase and decrease algorithms for congestion avoidance in computer networks,} {\protect\JournalTitle{Computer Networks and ISDN Systems}} \textbf{17}, 1--14 (1989). {\href{https://doi.org/10.1016/0169-7552(89)90019-6}{doi:10.1016/0169-7552(89)90019-6}}.

\bibitem{jacobson1988congestion}
V.~Jacobson, \enquote{Congestion avoidance and control,} {\protect\JournalTitle{ACM SIGCOMM Computer Communication Review}} \textbf{18}, 314--329 (1988). {\href{https://doi.org/10.1145/52324.52356}{doi:10.1145/52324.52356}}.

\end{thebibliography}


\begin{thebibliography}{1}
\newcommand{\enquote}[1]{``#1''}

\bibitem{rahimi2007random}
A.~Rahimi and B.~Recht, \enquote{Random features for large-scale kernel machines,} {\protect\JournalTitle{Advances in neural information processing systems}} \textbf{20} (2007).

\bibitem{schulman2015high}
J.~Schulman, P.~Moritz, S.~Levine, \emph{et~al.}, \enquote{High-dimensional continuous control using generalized advantage estimation,} {\protect\JournalTitle{arXiv preprint arXiv:1506.02438}}  (2015).

\end{thebibliography}






\end{document}


\providecommand{\theabstract}{}
\maketitle

\section{Extended Methods}
\label{sec:methods_supp}

\subsection{State Representation and Feature Encoding}
\label{subsec:state_repr_supp}

\begin{figure*}[t]
    \centering
    \includegraphics[width=\textwidth]{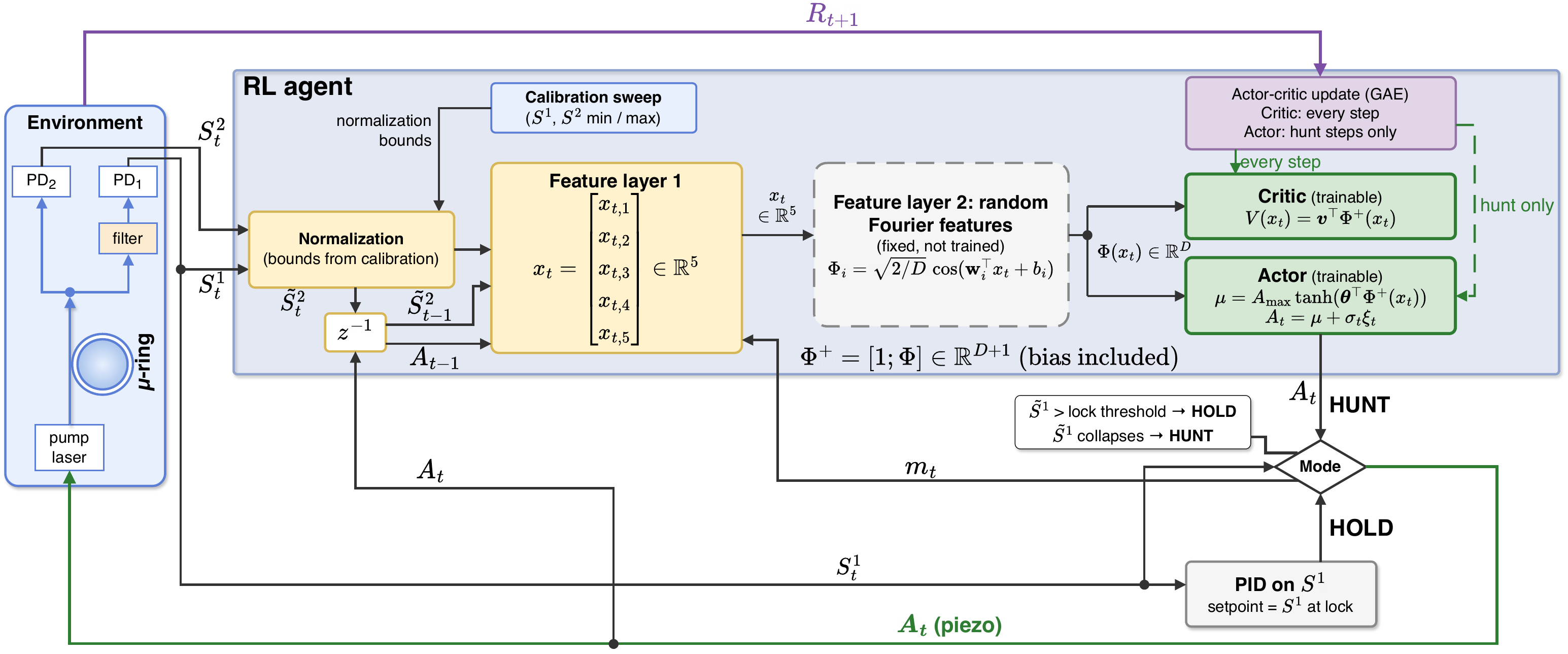}
    \caption{\textbf{Reinforcement learning framework and its hardware implementation.}
    The environment (blue) comprises the pump laser, the micro-ring resonator ($\mu$-ring), and two photodiodes measuring the filtered comb signal $S_t^1$ (PD$_1$) and the transmission signal $S_t^2$ (PD$_2$). The agent, implemented on the system-on-chip (SoC), normalizes these signals with bounds from an initial calibration sweep and assembles the state $\mathbf{x}_t\in\mathbb{R}^5$ (Feature layer~1, ~\eqref{eq:state}); $z^{-1}$ is a one-step delay. A fixed random Fourier feature map (Feature layer~2, ~\eqref{eq:rff}, dashed gray) feeds the trainable Critic and Actor (green), which are updated from the reward $R_t$ using generalized advantage estimation (Critic every step, Actor on hunt steps only). The Actor outputs the action $A_t$, a change in the piezo voltage applied to the pump laser. Once the comb is detected, the mode switch hands control to a PID loop on $S^1$; if the comb is lost, the hunt resumes. The mode flag $m_t$ is fed back as $x_{t,5}$. The loop period is 2.7~ms and the rollout length is $T=64$ (Table~\ref{tab:hyperparams}).}
    \label{fig:rl_framework_supp}
\end{figure*}

The control problem is cast as a Markov decision process (MDP) in which the agent is the controller running on the Red Pitaya system-on-chip (SoC) and the environment is the microring resonator together with its optical and electronic hardware (Fig.~\ref{fig:rl_framework_supp}). The control loop runs at discrete time steps $t$ of duration $\Delta t_{\rm step}=2.7$\,ms. At each step, the agent receives two photodiode measurements, $\mathbf{S}_t=(S_t^1,S_t^2)$. Here $S_t^1$ is the voltage of PD$_1$, which detects the band-pass-filtered comb light and is proportional to the filtered comb power, and $S_t^2$ is the voltage of PD$_2$, which detects the total transmitted light and is proportional to the transmission-port power. Both voltages are digitized by the 14-bit, 125\,MS/s analog inputs of the Red Pitaya with a decimation factor of 64, so that each sample is the average of 64 consecutive ADC samples (512\,ns). The most recent sample of each channel is read 2.5\,ms after the piezo command is applied. The intracavity state (detuning, intracavity energy, temperature) is not directly observable. The agent therefore works with a constructed state $x_t$, built from the current and previous measurements, the previous action, and the operating mode (Feature layer~1 in Fig.~\ref{fig:rl_framework_supp}).

\paragraph{Normalization and state construction.}
An initial deterministic calibration scan (as explained in Fig. 2b main text) maps the operational boundaries of the cavity, establishing the minimum and maximum values $S^i_{\min}$ and $S^i_{\max}$ of each channel $i\in\{1,2\}$. These bounds are measured once, not learned. In the Normalization block, the incoming signals are min-max normalized to $[-1,1]$,
\begin{equation}
    \tilde{S}_t^{\,i} = \frac{2\,(S_t^{\,i}-S^{\,i}_{\min})}{S^{\,i}_{\max}-S^{\,i}_{\min}} - 1 .
    \label{eq:norm}
\end{equation}
The agent state (Feature layer~1) is the five-component vector
\begin{equation}
    x_t = \big[\,x_{t,1},\,x_{t,2},\,x_{t,3},\,x_{t,4},\,x_{t,5}\,\big]^\top \in \mathbb{R}^5,
    \label{eq:state}
\end{equation}
with components
\begin{enumerate}
    \item $x_{t,1}=\tilde{S}_t^{2}$, the normalized transmission power;
    \item $x_{t,2}=\tilde{S}_t^{2}-\tilde{S}_{t-1}^{2}$, the step-by-step change in the normalized transmission, which encodes the direction and speed of the tuning trajectory. The previous value $\tilde{S}_{t-1}^{2}$ is supplied by the $z^{-1}$ delay block;
    \item $x_{t,3}=\tanh(\tilde{S}_t^{1})$, the normalized filtered comb power passed through a hyperbolic tangent. Once the soliton forms, the comb power can rise above the maximum recorded during calibration, so $\tilde{S}_t^{1}$ can leave $[-1,1]$. The $\tanh$ maps any such value smoothly back into a bounded range (a soft boundary), so out-of-range or noisy readings do not produce large, destabilizing inputs to the feature map;
    \item $x_{t,4}=A_{t-1}$, the action applied in the immediately preceding step, also supplied by the $z^{-1}$ block. It lets the agent condition on its own recent tuning command, which the photodiode signals alone do not reveal;
    \item $x_{t,5}=m_t\in\{-1,1\}$, a binary mode indicator set by the Mode block ($m_t=-1$: RL hunt; $m_t=1$: PID hold, Fig.~\ref{fig:rl_framework_supp}). The Critic is updated in both modes, so that it learns the value of the full visited state space, including the locked region, and thereby credits the hunt steps that lead to the lock. The Actor is updated only in hunt, since in hold the PID loop, not the policy, sets the voltage. The flag lets the Critic and the Actor distinguish states generated under these two controllers, which would otherwise look identical (Sec.~\ref{subsec:policy_selection_supp})
\end{enumerate}

\paragraph{Feature map.}
Throughout, vectors are bold lowercase, matrices are bold uppercase, and scalars are italic. To ensure computational efficiency for direct deployment on the programmable SoC, the Actor-Critic models avoid deep neural networks. Instead, the 5-dimensional state is projected into a $D=16$-dimensional non-linear feature space (Feature layer~2 in Fig.~\ref{fig:rl_framework_supp}) using Random Fourier Features (RFF),
\begin{equation}
    \boldsymbol{\Phi}(\mathbf{x}_t) = \sqrt{\frac{2}{D}}\,\cos\!\big(\mathbf{W}\mathbf{x}_t + \mathbf{b}\big)\in\mathbb{R}^{D},
    \label{eq:rff}
\end{equation}
where the random projection matrix $\mathbf{W}\in\mathbb{R}^{D\times 5}$ has entries drawn once at initialization from $\mathcal{N}(0,\sigma_{\mathrm{rff}}^2)$, and the random phase offset $\mathbf{b}\in\mathbb{R}^{D}$ is drawn from $\mathcal{U}(0,2\pi)$. For Gaussian $\mathbf{W}$, the expected inner product of two feature vectors is the Gaussian kernel $\exp\!\big(-\gamma\|\mathbf{x}-\mathbf{x}'\|^2\big)$ with $\sigma_{\mathrm{rff}}=\sqrt{2\gamma}$ \cite{rahimi2007random}. We use the unit-bandwidth kernel, $\gamma=1$, i.e., $\sigma_{\mathrm{rff}}=\sqrt{2}$ (length scale $\ell=1/\sqrt{2\gamma}\approx0.71$); this default was not tuned for the experiment. With $D=16$ the kernel is only coarsely approximated, so the features serve as a fixed random basis rather than an exact kernel expansion. Together, $\mathbf{W}$ and $\mathbf{b}$ are the \emph{random feature parameters}: they are \emph{fixed}, frozen after initialization, never trained, and shared by the Actor and the Critic (dashed gray block in Fig.~\ref{fig:rl_framework_supp}). We write $\boldsymbol{\Phi}^+(\mathbf{x}_t)=[1,\boldsymbol{\Phi}(\mathbf{x}_t)^\top]^\top\in\mathbb{R}^{D+1}$ for the feature vector augmented with a constant bias input.

\paragraph{Nonlinearity without a deep network.}
In a conventional neural network, nonlinear activation functions sit between trainable layers, so every layer must be trained by backpropagation. Here the nonlinearity is placed entirely in the fixed feature map of Eq.~\ref{eq:rff}. Each feature is a cosine of a random projection of the state, so the 16 features respond to different, random combinations and length scales of the state variables. The map is equivalent to a hidden layer with cosine activations whose connections ($\mathbf{W}$, $\mathbf{b}$) are random and frozen. For Gaussian $\mathbf{W}$, such features approximate a Gaussian (radial-basis) kernel, with the length scale set by $\sigma_{\mathrm{rff}}$ \cite{rahimi2007random}. Only the weights that combine these features, the critic weights $\mathbf{v}$ and the actor weights $\boldsymbol{\theta}$ defined below, are trained. Because the features are nonlinear in $\mathbf{x}_t$, a weighted sum of them can represent nonlinear functions of the state, such as the abrupt, state-dependent changes of the cavity response around the soliton step, while the learning problem stays simple: the quantities to be trained enter linearly.

The two trainable models are these weighted sums of the features. The Critic estimates the state value as
\begin{equation}
    V(\mathbf{x}_t) = \mathbf{v}^\top\boldsymbol{\Phi}^+(\mathbf{x}_t),
    \label{eq:critic_value}
\end{equation}
with the trainable \emph{critic weights} $\mathbf{v}\in\mathbb{R}^{D+1}$, while the Actor computes the mean action as
\begin{equation}
    \mu(\mathbf{x}_t) = A_{\max}\tanh\!\big(\boldsymbol{\theta}^\top\boldsymbol{\Phi}^+(\mathbf{x}_t)\big),
    \label{eq:actor_mean}
\end{equation}
with an independent set of trainable \emph{actor weights} $\boldsymbol{\theta}\in\mathbb{R}^{D+1}$. The outer $\tanh$ is a fixed output squashing that, together with the constant $A_{\max}$, bounds the mean action to $[-A_{\max},A_{\max}]$. $A_{\max}$ is fixed and calibrated to the physically safe tuning range of the piezoelectric actuator. In total, the only trainable parameters are $\mathbf{v}$ and $\boldsymbol{\theta}$ (green blocks in Fig.~\ref{fig:rl_framework_supp}), $2(D+1)=34$. The calibration bounds, $\mathbf{W}$, $\mathbf{b}$, $A_{\max}$, and the hyperparameters of Table~\ref{tab:hyperparams} are fixed.

\paragraph{Training cost.}
Since the models are weighted sums over a fixed feature map, each update requires only small vector operations, with no backpropagation through multiple layers. This removes most of the compute and memory cost of deep RL, which typically relies on GPUs and on offline or simulated training to absorb that cost. In our setting, learning runs directly on the embedded processor while the experiment is running, using measurements from the cavity taken during the experiment. The update computation must therefore be light enough to keep pace with the physical tuning process, which is what the linear-in-the-weights design provides. As a result, the amount of experimental time available for training, rather than model capacity, is the main practical constraint.

\subsection{Actor-Critic Reinforcement Learning Framework and Policy Update}
\label{subsec:actor_critic_supp}

\paragraph{MDP formulation.}
 At each discrete time step $t$, the agent builds the state $x_t\in\mathbb{R}^5$ from the photodiode signals (Feature layer~1; \eqref{eq:state}) and applies an action $A_t$, the change in the piezo tuning voltage. The environment then evolves under its unknown thermal and Kerr dynamics. It returns the next measurements, from which $x_{t+1}$ is built, and a scalar reward $R_t$ that quantifies the efficacy of the applied action. The reward is indexed by the action that produced it; the main text writes the same quantity as $R_{t+1}$. Table~\ref{tab:rl_mapping} maps each RL element to its physical realization.

\begin{table}[H]
\centering
\caption{Mapping between RL elements and the experiment (see Fig.~\ref{fig:rl_framework_supp}).}
\label{tab:rl_mapping}
\small
\begin{tabular}{@{}p{3.2cm}p{9.5cm}@{}}
\hline
RL element & Physical realization \\
\hline
Environment & Pump laser, $\mu$-ring, filter, PD$_1$, PD$_2$ (blue block) \\
Observation & $S_t=(S_t^1,S_t^2)$: filtered comb power and transmission power \\
State & $x_t\in\mathbb{R}^5$, built from $\tilde S_t$, $\tilde S_{t-1}$, $A_{t-1}$, $m_t$ (\eqref{eq:state}) \\
Action & $A_t$: change in piezo voltage applied to the pump laser (green $A_t$ line) \\
Reward & $R_t$: comb power plus shaping terms (\eqref{eq:reward}); arrow $R_t$ \\
Policy & Clipped Gaussian with mean from the Actor block (\eqref{eq:policy}) \\
End of hunt & Comb detected ($x_{t+1,3}\ge\tanh(2\rho-1)$): Mode block hands over to PID \\
End of episode & Comb lost ($x_{t,3}<x_{\mathrm{lost}}$), hunt-step limit $N_{\max}$ reached, or red end of the search window reached: a new hunt starts from the blue-detuned edge \\
\hline
\end{tabular}
\end{table}

One time step is one full pass of the control loop: application of the piezo command, a fixed delay of 2.5~ms for the PZT to settle (which avoids exciting its acoustic ringing), photodiode acquisition, normalization and state construction, and feature evaluation and action computation. The settling delay dominates the step duration. Together with about 0.2~ms of acquisition, processing and logging on the embedded processor, it gives a measured median step duration of $\Delta t_{\mathrm{step}}\approx2.7$~ms, the same in the hunt and hold phases. All time indices $t$, the rollout length $T$, and the exploration schedule are counted in these steps.

The cavity detuning, intracavity energy, and temperature are not measured directly, so the agent acts on the constructed state $x_t$ rather than on the true cavity state. The problem is therefore partially observed, and we treat it as an MDP on $x_t$. The previous action $A_{t-1}$ and the transmission change $x_{t,2}$ are included in $x_t$ to add some memory element.

\paragraph{Policy.}
In the hunt phase, the policy $\pi_\theta$ is Gaussian,
\begin{equation}
    A_t\sim\pi_\theta(\cdot\mid x_t)=\mathcal{N}\big(\mu(x_t),\,\sigma_t^2\big),
    \quad\text{i.e.}\quad
    A_t=\mu(x_t)+\sigma_t\,\xi_t,\ \ \xi_t\sim\mathcal{N}(0,1).
    \label{eq:policy}
\end{equation}
Here $\mu(x_t)$ is the mean action produced by the Actor (\eqref{eq:actor_mean}), whose trainable weight vector $\theta\in\mathbb{R}^{D+1}$ parametrizes the policy. $\sigma_t>0$ is the exploration standard deviation, annealed during each hunt (\eqref{eq:exploration_decay}), and $\xi_t$ is a unit-variance Gaussian noise sample drawn independently at each step. The term $\sigma_t\xi_t$ is the exploration noise added to the Actor output in Fig.~\ref{fig:rl_framework_supp}. The executed action is clipped to $[-A_{\max},A_{\max}]$, and the resulting tuning voltage is confined to a safe window set from the calibration sweep. $A_{\max}$ is a device-specific constant, set for each micro-ring to a safe per-step tuning range of the piezoelectric actuator. In the hold phase, the action $A_t$ is set by the PID loop instead of the policy, and no exploration noise is added.

\paragraph{Reward.}
The reward combines the comb power with two small shaping terms that encode prior physical knowledge to accelerate convergence,
\begin{equation}
    R_{t} = x_{t+1,3} + \mathcal{G}_{\mathrm{trans}}(x_{t+1,3}) + \mathcal{D}_{\mathrm{dir}}(A_t)
    \label{eq:reward}
\end{equation}
Here $x_{t+1,3}=\tanh(\tilde S^{1}_{t+1})$ is the soft-bounded normalized comb power measured after the action (~\eqref{eq:norm}), rescaled with the bounds from the calibration sweep. The calibration curve (Fig.~1(b) of the main text) shows that the soliton step is reached by tuning from the blue-detuned side toward the red-detuned side of the resonance. The two shaping terms encode only this known direction. $\mathcal{G}_{\mathrm{trans}}(\mathbf{x}_{t+1})$ is linear in the normalized transmission: it penalizes the high-transmission, blue-detuned region and rewards states close to resonance. $\mathcal{D}_{\mathrm{dir}}(A_t)$ depends only on the sign of the applied action: it rewards a step toward red detuning and penalizes a step toward blue detuning. Both terms have a fixed functional form and fixed coefficients, set before training, identical for all trials, and not learned. They only bias the agent toward the known direction of the soliton step and do not constrain it: the policy chooses both the direction and the size of every step, and can move away from this prior, for example by stepping back toward blue detuning, whenever this yields a higher return. Together, the two shaping terms contribute approximately $20\%$ of the reward, so the comb-power term dominates. Because the shaping terms are added to a comb term that lies in $(-1,1)$, the total reward is not confined to $[-1,1]$ (see the reward trace in Fig.~2(b) of the main text). The fine-tuning dynamics themselves are learned through direct interaction with the cavity, not from a pre-computed transition map or a fixed heuristic.

\paragraph{Objective.}
The agent seeks the policy that maximizes the expected discounted return,
\begin{equation}
    J(\theta)=\mathbb{E}_{\pi_\theta}\Big[\sum_{t=0}^{\infty}\gamma^{t}R_t\Big],
    \label{eq:objective}
\end{equation}
where $\gamma\in[0,1)$ is the discount factor that weights future rewards relative to immediate ones ($\gamma=0.99$; Table~\ref{tab:hyperparams}). No explicit model of the underlying optical physics is required.

\paragraph{Actor and Critic.}
We use an Actor-Critic architecture, which combines policy-gradient and value-based learning in two models that share the fixed feature map $\Phi^+(x_t)$ (green blocks in Fig.~\ref{fig:rl_framework_supp}). The Actor maps the state to a continuous control action through the policy of \eqref{eq:policy}. The Critic predicts the expected long-term reward from a state, $V(x_t)$ (\eqref{eq:critic_value}). Instead of updating the policy from raw, noisy rewards, the Actor is updated with an \emph{advantage} $\hat{\mathcal{A}}_t$: how much better the outcome of the action taken at step $t$ was than the Critic expected. The advantage is computed with Generalized Advantage Estimation (GAE, below) \cite{schulman2015high}. This reduces the variance of the policy-gradient estimate and gives faster, more stable convergence during real-time hardware training.

\paragraph{Two-phase operation and episodes.}\label{subsec:policy_selection_supp}
Operation alternates between two phases, set by the Mode block in Fig.~\ref{fig:rl_framework_supp}. An \emph{episode} begins with a \emph{hunt} ($m_t=-1$): the tuning voltage is set to an initial value, the exploration level is reset to $\sigma_0$, and the Actor samples actions from \eqref{eq:policy} while the agent explores. The hunt ends when the filtered comb power exceeds a fraction $\rho=0.90$ of its calibrated range. In normalized units this is $\tilde{S}^{1}_{t+1}\ge 2\rho-1=0.8$, which corresponds to $x_{t+1,3}\ge\tanh(0.8)\approx0.664$ after the $\tanh$ transform (Table~\ref{tab:hyperparams}). The Mode block then switches to the \emph{hold} phase ($m_t=1$) and records the filtered comb signal measured at that moment, after the short settling delay, as the PID setpoint $S^1_{\rm set}$. The threshold therefore only triggers the hold, whereas the setpoint is the measured value of $S^1$ at lock, which differs from trial to trial. In the hold phase, exploration is suspended, and a Proportional-Integral-Derivative (PID) loop takes authority over the tuning voltage, acting on the error $S^1_{\rm set}-S^1_t$ to suppress thermal drift and maintain the soliton.The episode ends in one of three ways. If the comb is lost during the hold ($x_{t,3}<x_{\mathrm{lost}}=-0.5$), a new hunt begins. If no comb is found within $N_{\max}$ hunt steps, or if the tuning voltage reaches the red-detuned end of the search window without a lock, the hunt is also terminated and a new one begins. In both cases the system is reset to the blue-detuned edge of the window rather than relying on the agent to learn to step back. Past the soliton step there is nothing left to find, so a stalled hunt would only generate uninformative data, and the reset costs no learning: it is more sample-efficient than waiting for exploration to return, or rewarding the agent for returning, to the blue side.

\paragraph{Exploration schedule.}
During the hunt phase, the exploration standard deviation decays geometrically with the number of hunt steps,
\begin{equation}
    \sigma_t=\max\!\big(\sigma_{\min},\ \sigma_0\,\kappa^{\,n_t}\big),
    \label{eq:exploration_decay}
\end{equation}
where $n_t$ is the number of hunt steps elapsed in the current episode, $\sigma_0=0.05$ is the initial exploration level, $\sigma_{\min}=0.005$ is the floor value, and $\kappa\in(0,1)$ is the per-step decay rate ($\kappa=0.9995$). The schedule is reset at the start of every hunt. It progressively shifts the policy from exploration toward exploitation as the hunt proceeds.

\paragraph{Rollouts.}
Learning is batched and on-policy. Transitions are collected over rollouts of at most $T=64$ steps ($\approx173$~ms at $\Delta t_{\mathrm{step}}\approx2.7$~ms). A rollout continues through the transition from hunt to hold. The weights are updated after every $T$ steps, or earlier if the episode ends. A rollout cut short by the end of an episode is treated as terminal, with zero bootstrap value. The complete loop, including the hunt-to-hold switch, the episode restart, and the call to the update routine, is given in Algorithm~\ref{alg:rollout}.

\begin{algorithm}[H]
\caption{On-Hardware Hunt/Hold Episodes with Rollout Collection}
\label{alg:rollout}
\begin{algorithmic}[1]
\Require Weights $\boldsymbol{\theta}$, $\mathbf{v}$; rollout length $T$; thresholds $\rho$, $x_{\mathrm{lost}}$; hunt-step limit $N_{\max}$
\Repeat \Comment{one pass = one episode}
    \State Reset piezo voltage; set mode to \textsc{hunt}; reset $\sigma$ to $\sigma_0$; clear buffer $\mathcal{D}$
    \State Measure $\mathbf{S}$ and build the state $\mathbf{x}$
    \Repeat \Comment{one pass = one control step}
        \If{mode is \textsc{hunt}}
            \State $A \leftarrow \mu(\mathbf{x}) + \sigma\,\xi$, with $\xi\sim\mathcal{N}(0,1)$, clipped to $\pm A_{\max}$
            \State Decay $\sigma$ (Eq.~\ref{eq:exploration_decay})
        \Else
            \State $\xi \leftarrow 0$; $A \leftarrow$ PID correction toward the setpoint
        \EndIf
        \State Apply $A$ to the piezo, wait for settling, measure $\mathbf{S}'$, build the next state $\mathbf{x}'$
        \State Compute reward $R$ (Eq.~\ref{eq:reward}) and store $(\mathbf{x},\xi,R)$ in $\mathcal{D}$
        \If{mode is \textsc{hunt} \textbf{and} comb detected ($x'_{3}\ge\tanh(2\rho-1)$)}
            \State Switch to \textsc{hold}; set PID setpoint to the current $S^1$
        \EndIf
        \State $\text{end} \leftarrow$ (\textsc{hold} and comb lost, $x'_{3}<x_{\mathrm{lost}}$) \textbf{or} (\textsc{hunt} and $N_{\max}$ steps used)
        \If{$|\mathcal{D}|=T$ \textbf{or} end}
            \State Update $\boldsymbol{\theta},\mathbf{v}$ from $\mathcal{D}$ (Alg.~\ref{alg:update}), with bootstrap value $0$ if end, else $V(\mathbf{x}')$
            \State Clear $\mathcal{D}$
        \EndIf
        \State $\mathbf{x}\leftarrow\mathbf{x}'$
    \Until{end}
\Until{stopped by the operator}
\end{algorithmic}
\end{algorithm}

\paragraph{Advantage estimation.}
The Critic's one-step temporal-difference (TD) error,
\begin{equation}
    \delta_t = R_t + \gamma\,V(x_{t+1}) - V(x_t),
    \label{eq:gae_delta}
\end{equation}
compares the reward plus the discounted value of the next state with the value predicted for the current state. GAE combines the TD errors over the remainder of the rollout of length $L\le T$,
\begin{equation}
    \hat{\mathcal{A}}_t=\sum_{k=0}^{L-t-1}(\gamma\lambda)^{k}\,\delta_{t+k},
    \label{eq:gae_advantage}
\end{equation}
where $\lambda\in[0,1]$ is the GAE decay parameter. Small $\lambda$ gives low-variance but biased estimates, and large $\lambda$ the reverse. We use $\gamma=0.99$ and $\lambda=0.95$ to balance bias and variance. The value target for the Critic is $\hat{V}_t=\hat{\mathcal{A}}_t+V(x_t)$.

\paragraph{Weight updates.}
Only the weight vectors $v$ (Critic) and $\theta$ (Actor) are trained (green blocks in Fig.~\ref{fig:rl_framework_supp}), directly on the embedded processor. Both are initialized to zero before the first hunt, so no pretraining is used; only the fixed feature map $(W,b)$ is random. At the start of each update, the Critic weights are shrunk multiplicatively,
\begin{equation}
    v\leftarrow \eta\,v,
    \label{eq:critic_decay}
\end{equation}
where $\eta\in(0,1)$ is the weight-decay factor ($\eta=0.999$). This is applied once per rollout, not per transition, to prevent unbounded growth of the value estimate. The Critic is then updated at every step of the rollout, whether the step occurred under hunting or under PID hold (solid green arrow in Fig.~\ref{fig:rl_framework_supp}),
\begin{equation}
    v\leftarrow v+\alpha_{\mathrm{critic}}\big(\hat{V}_t-V(x_t)\big)\,\Phi^{+}(x_t),
    \label{eq:critic_update}
\end{equation}
with Critic learning rate $\alpha_{\mathrm{critic}}=0.005$. The Actor is updated only on steps where the action was sampled from the stochastic hunt policy (dashed green arrow in Fig.~\ref{fig:rl_framework_supp}),
\begin{equation}
    \theta\leftarrow\theta+\alpha_{\mathrm{actor}}\,\hat{\mathcal{A}}_t\,\xi_t\,\Phi^{+}(x_t)
    \qquad(\text{hunt steps only, }\xi_t\neq0),
    \label{eq:actor_update}
\end{equation}
with Actor learning rate $\alpha_{\mathrm{actor}}=0.001$. Here the noise sample $\xi_t$ of \eqref{eq:policy} serves as a lightweight proxy for the policy score function, the gradient of $\ln\pi_\theta$ with respect to $\theta$. The exact score of the Gaussian policy would contain the additional factor $A_{\max}\big(1-\tanh^2(\theta^\top\Phi^+(x_t))\big)/\sigma_t$, which the proxy omits. This avoids evaluating the full score on embedded hardware, at the cost of an approximate rather than exact policy-gradient estimator. Steps executed under PID hold carry no exploration noise ($\xi_t=0$), so they are excluded from the Actor update but still contribute to the Critic update. Algorithm~\ref{alg:update} summarizes the full update.

\begin{algorithm}[H]
\caption{On-Policy Actor-Critic Update via GAE}
\label{alg:update}
\begin{algorithmic}[1]
\Require Buffer $\{(x_t,\xi_t,R_t,\Phi^+(x_t))\}_{t=0}^{L-1}$ with $L\le T$, weights $\theta$, $v$, parameters $\gamma$, $\lambda$, $\alpha_{\mathrm{actor}}$, $\alpha_{\mathrm{critic}}$, $\eta$, bootstrap value $V_{\mathrm{boot}}$ (used as $V(x_L)$)
\For{$t=L-1,\dots,0$} \Comment{backward GAE recursion}
    \State $\delta_t\leftarrow R_t+\gamma V(x_{t+1})-V(x_t)$
    \State $\hat{\mathcal{A}}_t\leftarrow\delta_t+\gamma\lambda\,\hat{\mathcal{A}}_{t+1}$ \Comment{$\hat{\mathcal{A}}_L:=0$}
    \State $\hat{V}_t\leftarrow\hat{\mathcal{A}}_t+V(x_t)$
\EndFor
\State $v\leftarrow\eta\,v$ \Comment{Critic weight decay, once per rollout}
\For{$t=0,\dots,L-1$}
    \State $v\leftarrow v+\alpha_{\mathrm{critic}}\big(\hat{V}_t-V(x_t)\big)\Phi^+(x_t)$ \Comment{Critic: every step}
    \If{$\xi_t\neq0$} \Comment{Actor: hunt steps only}
        \State $\theta\leftarrow\theta+\alpha_{\mathrm{actor}}\,\hat{\mathcal{A}}_t\,\xi_t\,\Phi^+(x_t)$
    \EndIf
\EndFor
\State \Return updated $\theta$, $v$
\end{algorithmic}
\end{algorithm}

Table~\ref{tab:hyperparams} collects all hyperparameters used in the above.

\begin{table}[H]
\centering
\caption{Actor-Critic hyperparameters. All constants are fixed at compile time and identical across devices, except $A_{\max}$, which is device-specific ($\dagger$): it is set for each micro-ring to a safe per-step tuning range of the piezoelectric actuator.}
\label{tab:hyperparams}
\small
\begin{tabular}{@{}l p{8.5cm} l@{}}
\hline
Symbol & Meaning & Value \\
\hline
$D$ & RFF feature dimension & 16 \\
$\sigma_{\mathrm{rff}}$ & Std.\ of RFF matrix entries $W_{ij}$ (sets feature length scale) & $\sqrt{2}$ \\
$T$ & Rollout horizon (steps per update) & 64 \\
$\gamma$ & Discount factor & 0.99 \\
$\lambda$ & GAE decay parameter & 0.95 \\
$\alpha_{\mathrm{actor}}$ & Actor learning rate & 0.001 \\
$\alpha_{\mathrm{critic}}$ & Critic learning rate & 0.005 \\
$\eta$ & Critic weight-decay factor (per rollout) & 0.999 \\
$\sigma_0$ & Initial exploration std. & 0.05 \\
$\sigma_{\min}$ & Final exploration std. & 0.005 \\
$\kappa$ & Exploration decay rate (per hunt step) & 0.9995 \\
$N_{\max}$ & Hunt-step limit per episode & 5000 \\
$x_{\mathrm{lost}}$ & Comb-loss threshold on $x_{t,3}$ (hold $\to$ hunt) & $-0.5$ \\
 & PZT settling delay after each piezo command & 2.5~ms \\
$\Delta t_{\mathrm{step}}$ & Duration of one control step (settling delay, acquisition, processing, actuation), median & $\approx2.7$~ms \\
$\rho$ & Comb-detection threshold, fraction of calibrated range & 0.90 \\
 & \quad equivalent value of $x_{t,3}$ ($\tanh$ space) & $\tanh(2\rho-1)\approx0.664$ \\
\hline
\end{tabular}
\end{table}

\clearpage

\section{Failure Mode Analysis}
\label{sec:failure_mode_supp}

\begin{figure}[!htbp]
    \centering
    \includegraphics[width=0.95\linewidth]{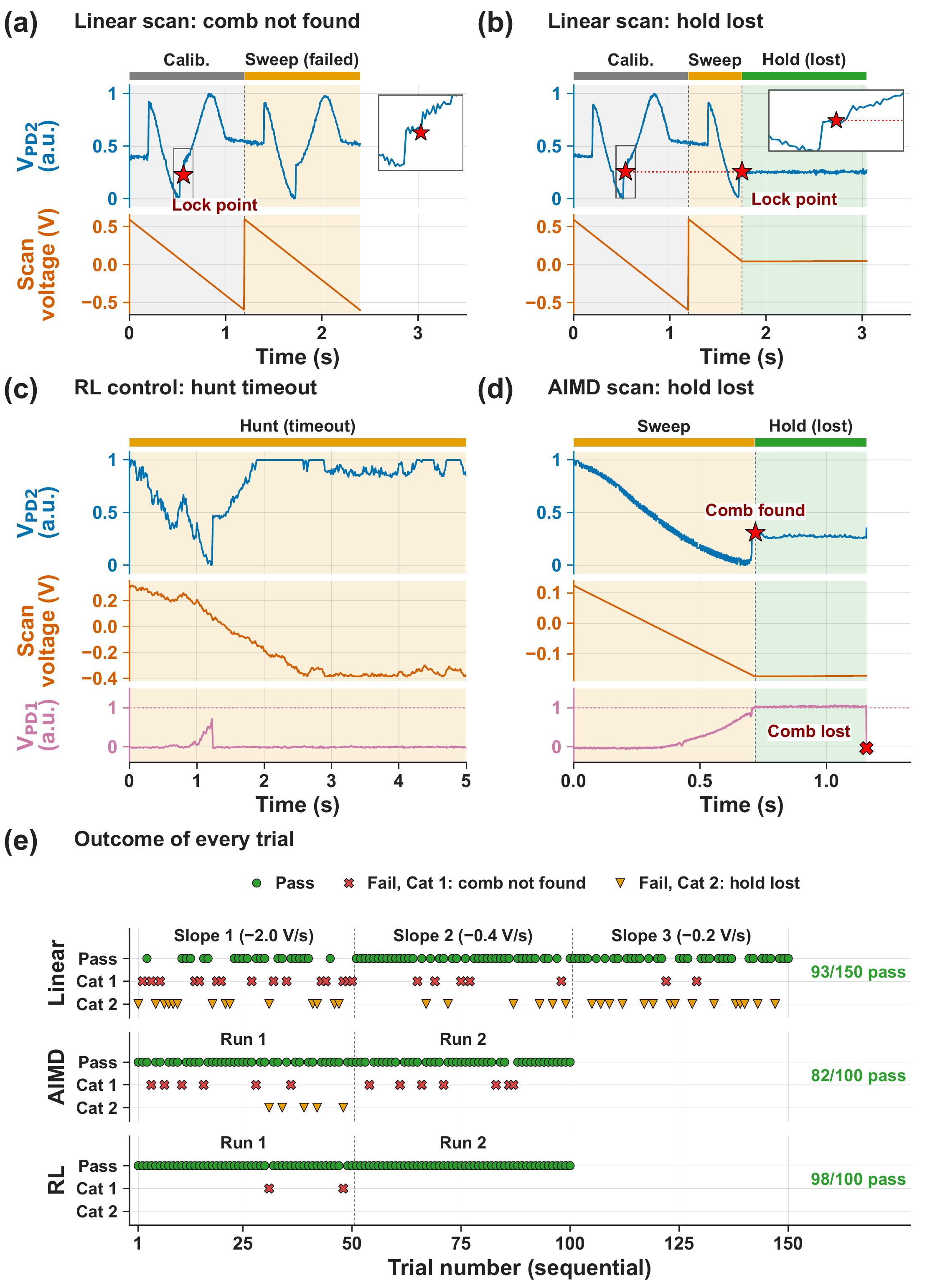}
    \caption{
    Failure mode analysis of fixed-slope, AIMD, and RL control.
    \textbf{(a)} Fixed-slope linear scan, Category~1 (comb not found): the sweep passes through the soliton step without triggering the hold.
    \textbf{(b)} Fixed-slope linear scan, Category~2 (hold lost): the hold engages on an unstable edge of the soliton step and the comb collapses to the CW state.
    \textbf{(c)} RL control, Category~1: the agent never reaches the soliton step, so no hold is triggered, and the hunt ends when the tuning range is exhausted.
    \textbf{(d)} AIMD linear scan, Category~2: the scan locks onto the soliton step but subsequently loses the hold.
    In (a)--(d), the phase of each trial is labeled above the panel, and stars mark the lock point or the point where the comb is found.
    \textbf{(e)} Outcome of every trial, with pass, Category~1, and Category~2 outcomes marked by different symbols. Fixed-slope failures follow a slope-dependent pattern (22 Category~1, 35 Category~2), AIMD Category~2 failures (5 of 18) cluster within its first evaluation campaign before disappearing entirely, and the two RL failures (both Category~1) are isolated and recover autonomously.
    }
    \label{fig:failure_modes}
\end{figure}

To identify the mechanisms behind the lower success rates of the linear-scan baselines, we analyzed the failure dynamics of the same trials as in the main text: three fixed-slope linear scans of 50 trials each (150 trials), the AIMD linear scan (100 trials, two campaigns of 50), and the RL agent (100 trials, two campaigns of 50), the latter starting from randomly initialized policy weights with no pretraining. Fig.~\ref{fig:failure_modes} shows representative failure events for each method together with trial-by-trial timelines of all outcomes, which reveal clearly different failure behavior for the fixed-slope, AIMD, and RL approaches.
All failures of the fixed-slope and AIMD baselines fall into one of two archetypes, both consistent with an inability to fully compensate for real-time cavity dynamics and thermal drift:

\begin{itemize}
    \item \textbf{Category 1 Failure (Comb Not Found):} As shown for the fixed-slope linear scan in Fig.~\ref{fig:failure_modes}(a), the initial calibration pass successfully maps the cavity resonance and locates the soliton step. During the subsequent sweep, however, the voltage ramp passes through the resonance without triggering the hold phase. We interpret this as the sweep crossing the narrow detuning window too quickly to trigger a lock, consistent with this failure mode being most prevalent at the fastest scan speed tested.

    \item \textbf{Category 2 Failure (Hold Lost):} Fig.~\ref{fig:failure_modes}(b) shows a second failure type for the fixed-slope linear scan. Here the sweep reaches the soliton step and triggers the hold phase, but the soliton is lost shortly afterwards and the system collapses to the continuous-wave (CW) state. We suspect this occurs because the resonance drifts during the (slower) sweep, so that the hold engages on the vulnerable edge of the soliton step rather than near its center, from which the soliton cannot be sustained. We note that, since the absolute cavity resonance position is not independently tracked, this interpretation is inferred from the transmission and power traces rather than confirmed directly. Fig.~\ref{fig:failure_modes}(d) shows the same failure type in the AIMD linear scan: the ramp locates the soliton step and engages the hold, but the hold is subsequently lost before the 10~s success criterion is reached.
\end{itemize}

The RL agent's rare failures are of Category 1, meaning the hunt does not reach the soliton step and no hold is triggered. Unlike the baselines, these failures are not caused by a fixed sweep speed. Fig.~\ref{fig:failure_modes}(c) illustrates one such hunt-phase timeout. The dynamically computed trajectory fails to satisfy the conditions required to reach the soliton step, the cavity transmission recovers to baseline, and the agent exhausts its available tuning range without securing a lock. No RL trial showed a Category 2 failure, in which the lock is engaged and then lost.

The aggregated timelines in Fig.~\ref{fig:failure_modes}(e) highlight the most critical operational differences between the three frameworks. The fixed-slope scan's failure mode is dictated by its scan speed, with each speed regime exhibiting a distinct, recurring failure signature rather than a progressive worsening over time. The AIMD linear scan's failures show an intermediate pattern: Category 1 misses persist at a comparable rate throughout both evaluation campaigns, whereas Category 2 failures are concentrated entirely within the first campaign and are absent from the second. The RL agent's rare failures, by contrast, are isolated events, after each of which the agent succeeds on the next trial.

\textbf{Slope-Dependent Failures (Fixed-Slope Scans):} Rather than compounding uniformly over the full trial sequence, the fixed-slope scans' failures appear to be dominated by a different mechanism within each slope regime. At Slope~1 ($-2.0\,\text{V/s}$, the fastest scan), Category 1 (comb not found) misses dominate, which we interpret as the rapid sweep spending too little time within the detuning window to trigger a lock before passing through it. At Slope~2 ($-0.4\,\text{V/s}$), both failure categories are comparatively sparse, consistent with this slope representing the best-performing empirical compromise (78\%). At Slope~3 ($-0.2\,\text{V/s}$, the slowest scan), Category 1 misses are rare, but Category 2 (hold lost) failures recur densely throughout the block, which we interpret as the slower sweep reliably reaching the soliton step, only for accumulated thermal drift to have shifted the resonance so that the hold engages on the unstable edge of the step. Across the full $N = 150$ fixed-slope trials, this yields 22 Category 1 and 35 Category 2 failures in total, each concentrated within the slope regime where its underlying mechanism appears to dominate. We note that these interpretations are based on the observed transmission and power traces rather than an independent measurement of the cavity's absolute resonance position, which was not tracked in this experiment. Slope~2 thus represents an empirically favorable, yet fragile, manually tuned compromise between these two competing failure modes, a balance the RL agent does not need to strike by hand because it adapts its trajectory in response to live cavity feedback.

\begin{figure}[!htbp]
\centering
\includegraphics[width=0.85\linewidth]{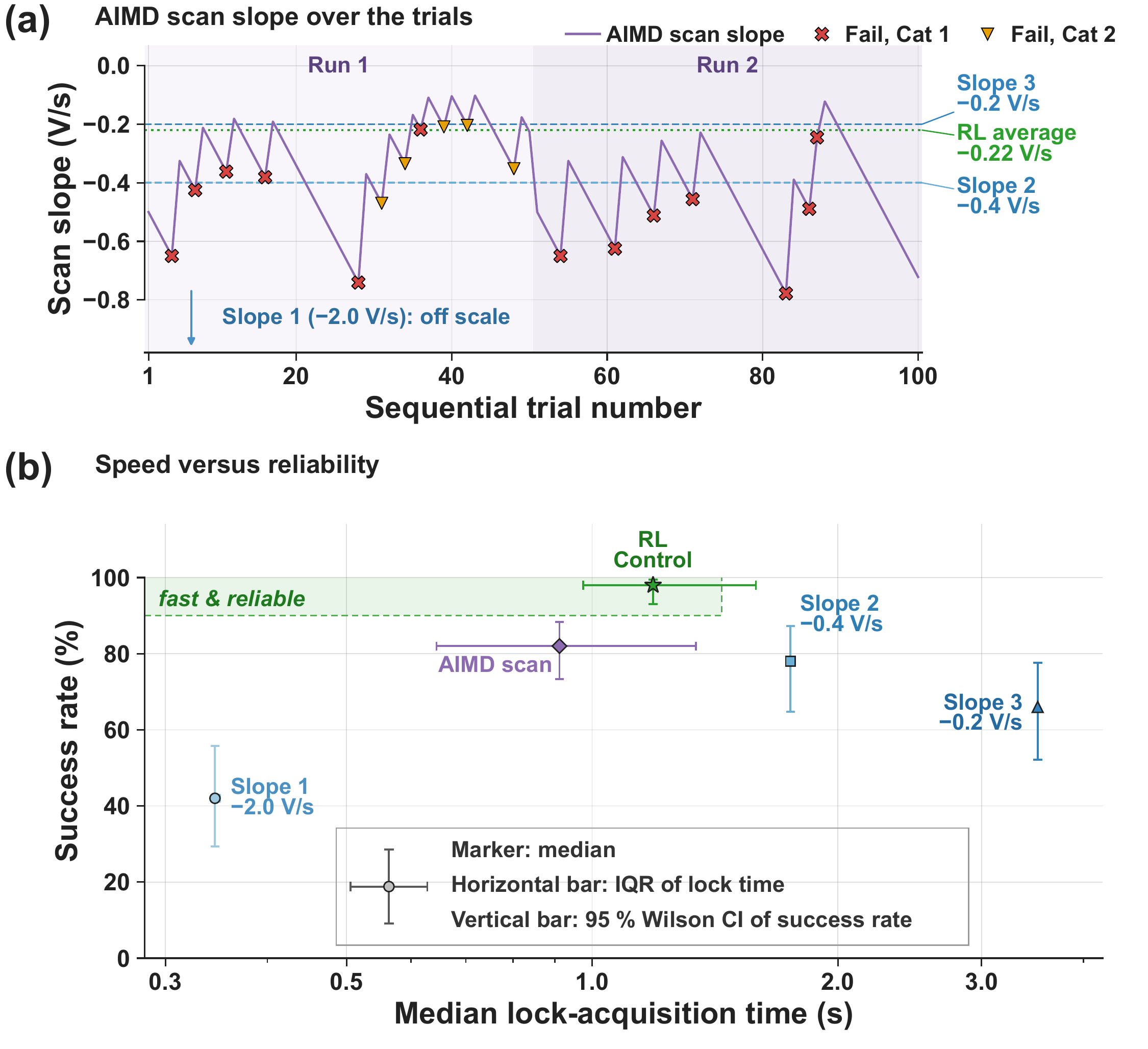}
\caption{
AIMD linear-scan speed dynamics and the speed--reliability trade-off.
\textbf{(a)} Scan slope of the AIMD linear scan (Eq.~\ref{eq:aimd_slope}) versus sequential trial number across both campaigns. Failed trials are marked by category (Category~1 or Category~2, as in Fig.~\ref{fig:failure_modes}) and all other trials passed. Dashed lines mark Slopes~2 and~3, the dotted line marks the RL agent's average sweep rate, and Slope~1 ($-2.0\,\text{V/s}$) lies outside the plotted range.
\textbf{(b)} Success rate versus median lock-acquisition time (logarithmic axis) for all five strategies. Markers show the success rate (vertical) and the median lock time (horizontal), vertical bars the 95\% Wilson score interval, and horizontal bars the interquartile range (IQR) of the lock time.
}
\label{fig:aimd_slope}
\end{figure}

\textbf{AIMD Scan-Slope Derivation:} Because the AIMD linear scan has no single fixed slope, relating its failures to the fixed-slope scans' slope-dependent mechanisms above requires an equivalent, trial-resolved measure of scan speed. For each AIMD trial, the logged step size $\Delta V_{\text{step}}$ (recorded once at the start of the trial) is converted to an instantaneous scan slope via
\begin{equation}
    \text{Slope} = -\frac{\Delta V_{\text{step}}}{\delta t},
    \label{eq:aimd_slope}
\end{equation}
where $\delta t$ is the fixed control-loop period between successive detuning updates during the hunt phase. This period was not assumed but measured directly from the logged timestamps: the median time between consecutive hunt-phase samples across both evaluation campaigns is $1.0\,\text{ms}$, consistent with the fixed control-loop delay set in firmware. In our implementation the step size starts at $5\times10^{-4}$~V per update, increases by $5\times10^{-5}$~V after each successful trial, is halved after each failed trial, and is clamped to $[1\times10^{-4},\,1.5\times10^{-3}]$~V, which at $\delta t = 1$~ms corresponds to slopes between 0.1 and 1.5~V/s.
The negative sign matches the sign convention used for the fixed-slope scans above. Fig.~\ref{fig:aimd_slope}(a) plots this derived slope against the sequential trial number across both AIMD campaigns, alongside the fixed-slope scans' slopes and the RL agent's average sweep rate ($-0.22\,\text{V/s}$) for reference, with failed trials marked by category (all other trials passed). This visualization makes the additive-increase, multiplicative-decrease behavior directly observable: the scan speed grows gradually (the slope becomes more negative) across consecutive successful trials, then drops sharply, as the step size is halved, immediately following any failure, producing the characteristic sawtooth trajectory.

\textbf{Failures of the AIMD Linear Scan:} The AIMD linear scan, evaluated over the same two-campaign protocol as the RL agent ($N=100$ trials total), fails on 18 of 100 trials (13 Category 1, 5 Category 2), a lower failure rate than any of the fixed slopes (58\%, 22\%, and 34\% for Slopes 1, 2, and 3). Unlike the fixed-slope scans, however, the two failure categories are not partitioned by a fixed operating regime chosen in advance, since AIMD has no such fixed regime; instead, they separate both temporally across the two evaluation campaigns and, as Fig.~\ref{fig:aimd_slope}(a) shows, by the scan slope active at the time of failure. All five Category 2 (hold lost) failures occur within the first campaign, clustered around trials 31 through 48 of that run, and none occur in the second campaign. Category 1 (comb not found) misses, by contrast, occur at a comparable rate in both campaigns (6 of 50 in the first, 7 of 50 in the second) and show no such temporal concentration.

These two failure categories also occupy distinct regions of the slope trajectory in Fig.~\ref{fig:aimd_slope}(a). Category 1 failures occur predominantly at faster scan slopes (mean $-0.50\,\text{V/s}$, range $-0.22$ to $-0.78\,\text{V/s}$) than the overall population of successful trials (mean $-0.38\,\text{V/s}$), consistent with the same mechanism identified for the fixed-slope baselines: a sweep moving too quickly is more likely to pass through the narrow detuning window without triggering a lock. Category 2 failures, in contrast, cluster at slower slopes (mean $-0.31\,\text{V/s}$, range $-0.20$ to $-0.47\,\text{V/s}$), sitting almost entirely within the band bracketed by Slope~2 and Slope~3/the RL average, the same speed regime in which the fixed-slope Slope~3 baseline exhibited its own Category 2 failures above. This is consistent with our interpretation that hold-lost failures arise not from excessive scan speed but from the hold engaging on the unstable edge of the soliton step after the resonance has drifted during the (comparatively slow, careful) scan.

We interpret the temporal clustering of Category 2 failures within the first campaign as reflecting the different timescales over which the two AIMD tuning mechanisms act: the additive step-size increase and window re-centering respond only after a successful trial, so a transient period of miscalibrated step size or drift compensation can persist across several consecutive trials before converging, plausibly accounting for the clustered run of Category 2 failures early in the first campaign, visible in Fig.~\ref{fig:aimd_slope}(a) as the slower, more erratic segment of the trajectory between trials 31 and 48. Once converged, this source of failure does not reappear, consistent with its complete absence from the second campaign. Category 1 misses, in contrast, appear to arise from instantaneous overshoot or undershoot events within an otherwise well-converged scan and are not resolved by the same convergence process, since they persist at a similar rate, and across a similarly broad range of slopes, throughout both campaigns. This suggests that AIMD's trial-to-trial tuning is effective at correcting persistent, systematic drift but does not protect against isolated single-trial excursions, a limitation the RL agent's within-trial adaptivity does not share.

\textbf{Isolated Failures (RL Control):} In contrast to both baselines, the RL agent's failures are exceptionally rare (2 out of 100 consecutive trials) and do not form a run of consecutive failures. These timeout events appear to be isolated stochastic anomalies rather than symptoms of systemic drift or an unconverged adaptation process. Because the RL agent continuously evaluates the measured physical state rather than relying on a historical calibration map or a slowly converging trial-to-trial update rule, a single failure does not precipitate cascading degradation, nor does it require a multi-trial burn-in period analogous to AIMD's Category 2 failures. The agent succeeds in the trials immediately following each missed event, without any operator intervention. We note, however, that both RL failures fall within the same part of the first campaign in which AIMD's Category 2 failures cluster (trials 31 through 48), so a transient environmental disturbance during that interval cannot be excluded.

\textbf{Overall Speed-Reliability Trade-off (Panel b):} Fig.~\ref{fig:aimd_slope}(b) situates all five strategies in a single trade-off space defined by success rate and median lock-acquisition time. The RL agent is the only strategy above 90\% success, with a median acquisition time shorter than those of Slopes~2 and~3 but longer than those of Slope~1 and AIMD. Among the baselines, no single fixed slope dominates: the fastest (Slope~1) sacrifices reliability for speed, the slowest (Slope~3) sacrifices both speed and reliability, and the intermediate slope (Slope~2) represents the best fixed-slope compromise, which AIMD improves on by tuning its speed from trial to trial, though still without reaching the RL agent's success rate.

\section{Statistical Methodology: Wilson Score Confidence Interval}
\label{subsec:wilson_supp}
To quantify the precision of the binomial success rates (pass/fail outcomes) across the linear, AIMD, and reinforcement learning trials, we compute the 95\% Wilson score interval, which is obtained by inverting the score test for a binomial proportion. Given $n$ total trials and $x$ successful outcomes, the sample proportion is $\hat{p} = x/n$. The lower and upper bounds of the confidence interval are calculated as:
\begin{equation}
    \text{CI}_{\pm} = \frac{\hat{p} + \frac{z^2}{2n} \pm z \sqrt{\frac{\hat{p}(1-\hat{p})}{n} + \frac{z^2}{4n^2}}}{1 + \frac{z^2}{n}}
    \label{eq:wilson_supp}
\end{equation}
where $z = 1.96$ corresponds to the critical value for a 95\% confidence level. Because this formulation accounts for the sample size directly in the denominator adjustment, it produces reliable, asymmetric error boundaries that accurately reflect statistical uncertainty. Table~\ref{tab:wilson} lists the resulting intervals for all five strategies; the lower bound for the RL agent (93.0\%) exceeds the upper bound of every baseline (88.3\% at most).

\begin{table}[!htb]
\centering
\caption{Success rates and 95\% Wilson score intervals (Eq.~\ref{eq:wilson_supp}) for all five strategies.}
\label{tab:wilson}
\begin{tabular}{lccc}
\hline
Strategy & $x/n$ & Success rate & 95\% Wilson interval \\
\hline
Slope 1 ($-2.0$~V/s) & 21/50 & 42.0\% & [29.4\%, 55.8\%] \\
Slope 2 ($-0.4$~V/s) & 39/50 & 78.0\% & [64.8\%, 87.2\%] \\
Slope 3 ($-0.2$~V/s) & 33/50 & 66.0\% & [52.2\%, 77.6\%] \\
AIMD & 82/100 & 82.0\% & [73.3\%, 88.3\%] \\
RL & 98/100 & 98.0\% & [93.0\%, 99.4\%] \\
\hline
\end{tabular}
\end{table}

\bibliography{sample_supp}